\documentclass[hyper]{JHEP3}
\pdfoutput=1
\usepackage{amsmath,amssymb,amsfonts,amsxtra,mathrsfs,makeidx,graphics,graphicx,amsthm}
\usepackage{epsfig}
\usepackage[all]{xy}

\def\IC{\mathbb{C}}

\def\R{\mathbb{R}}
\def\Z{\mathbb{Z}}
\def\Tr{{ \rm Tr}}
\def\a{\alpha}
\def\b{\beta}

\def\be{\dot{\beta}}

\def\p{\partial}
\def\be{\begin{equation}}
\def\ee{\end{equation}}
\def\bea{\begin{eqnarray}}
\def\eea{\end{eqnarray}}
\def\N{{\cal N}}
\def\S{{\cal S}}
\def\e{\epsilon}
\newcommand{\ndt}{\noindent}

\newcommand{\nslash}{\ensuremath \raisebox{0.025cm}{\slash}\hspace{-0.25cm} \nabla}
\newcommand{\Dslash}{\ensuremath \raisebox{0.025cm}{\slash}\hspace{-0.28cm} D}

\preprint{TIFR/TH/10-38}

\title{ Localization of ${\cal N}=4$ Superconformal Field Theory \\on $S^1 \times S^3$ and Index}

\author{Satoshi Nawata\\
Tata Insitute of Fundamental Research, \\
Mumbai India, 400005
\\
{\tt \email{ snawata@theory.tifr.res.in}} }

\abstract{We provide the geometrical meaning of the ${\cal N}=4$ superconformal index. With this interpretation, the ${\cal N}=4$ superconformal index can be realized as the partition function on a Scherk-Schwarz deformed background. We apply the localization method in TQFT to compute the deformed partition function since the deformed action can be written as a  $\delta_\epsilon$-exact form. The critical points of the deformed action turn out to be the space of flat connections which are, in fact, zero modes of the gauge field. The one-loop evaluation over the space of flat connections reduces to the matrix integral by which the ${\cal N}=4$ superconformal index is expressed. }
\begin{document}

\section{Introduction}

For the last few decades, there has been fruitful interaction between
quantum physics and geometry. This was triggered by the pioneering
work of Witten on supersymmetry and Morse theory \cite{Witten:1982im}
in which it was shown that the $0+1$-dimensional supersymmetric non-linear
sigma model with target a compact manifold $M$ (supersymmetric quantum
mechanics on $M$) is the de Rham-Hodge theory on $M$, and the Witten
index ${\rm Tr}\left(-1\right)^{F}e^{-\beta H}$ gives the Euler characteristic
$\chi(M)$ of the target manifold $M$. The paper \cite{Witten:1982im}
paved the way to study supersymmetric quantum field theory as de Rham-Hodge
theory of infinite dimensional manifolds \cite{Witten:1988xj,Witten:1988ze,Witten:1988hf}.

Quantum field theory has developed methods to deal with infinitely
many degrees of freedom based on Feynman functional (path) integral.
These methods are applied to extract a finite dimensional object out
of an infinite dimensional one by constructing topological invariants
as partition functions of fields on manifolds. Such quantum field
theory is in general called topological quantum field theory (TQFT) and can be classified as being of either of two types: Schwarz type or cohomological
(Witten) type. TQFTs of Schwarz type has a metric independent
classical action which is not a total derivative. It was heuristically
outlined in \cite{Witten:1988hf} that invariants of three-manifolds
and links in three-manifolds can be obtained as quantum Hilbert spaces
for the partition function of the Chern-Simons action, generalizing
the Jones polynomials \cite{Jones:1985dw,Jones:1987dy}. The constructions of \cite{Witten:1988hf} shed
new light, in particular, on the connection between three dimensional
topology and two-dimensional conformal field theory, and led to rigorous
definitions of the invariants in mathematics \cite{Reshetikhin:1991tc,Crane:1991jv,Kohno:1992,Kohno:1992hv}\footnote{In \cite{Reshetikhin:1991tc}, the invariants are expressed on the basis of the theory of quantum groups at roots of unity. In \cite{Crane:1991jv,Kohno:1992,Kohno:1992hv}, the invariants are constructed by the action of mapping class groups on the space of conformal blocks in two dimensional conformal field theory. It turns out that both the definitions are equivalent \cite{Piunikhin:1994}. }. On the other hand, an action of cohomological type depends on a metric, but inherits BRST-like symmetry $Q$ which is usually obtained by twisting supersymmety. The stress-energy tensor of TQFT can be written as $Q$-exact form, which implies that the vacuum expectation values of $Q$-invariant operators are 
independent of a metric, {\it i.e.}, the theory is topological. Although the precise mathematical definition of Feynman functional integral is not yet known, the $Q$-symmetry localizes Feynman functional integral to a finite dimensional integral over a certain moduli space, providing topological invariants. The realization, by quantum field theory, of the Gromov-Witten \cite{Witten:1988xj}, Donaldson-Witten \cite{Witten:1988ze} and Seiberg-Witten theory  \cite{Witten:1994cg}, and a strong coupling test of the $S$-duality carried out in \cite{Vafa:1994tf} can be seen as salient examples of TQFTs of cohomological type. 

Unlike TQFT of cohomological type, actions and stress-energy tensors of superconformal field theories (SCFTs) in four dimensions cannot be written as $Q$-exact form in general. Moreover, although a fermionic generator $\e$ for a BRST-like charge $Q$ can be regarded a scalar and can be set to be a non-zero constant  everywhere in TQFT, superconformal generators $\e_\a$ depend on the coordinates of a base manifold since they are solutions of conformal Killing spinor equations
\be
\nabla_\mu \e =-\frac 14 \gamma_\mu\nslash \e \ .
\label{ckse}
\ee
Therefore, one cannot simply apply the localization method in TQFT of cohomological type to compute partition functions of SCFTs exactly. However, motivated by the equivariant localization \cite{Duistermaat:1982vw,Atiyah:1982fa,Berline:1982,Atiyah:1984px,Witten:1992xu}, Pestun obtained exact results in \cite{Pestun:2007rz} for  the $\N=4$ SCFT on $S^4$ as well as the $\N=2$ and the $\N=2^*$ supersymmetric Yang-Mills theories (SYM) on $S^4$ by adding $\delta_\e$-exact term to the action. Here $\delta_\e$ is a fermionic symmetry generated by a suitable conformal Killing spinor $\e$. The Feymann functional integral of the $\N=4$ SCFT on $S^4$ is localized over the constant modes of the scalar field with all other fields vanishing. In this way, the vacuum expectation value of a supersymmetric Wilson line is also computed exactly.

 Following this example, we apply the method of the localization to the $\N=4$ SCFT on $S^1\times S^3$. Since the superconformal index is independent of the coupling constant, the action itself is presumably written as a $\delta_\e$-exact form in the case of $S^1\times S^3$. We will show that SCFTs on $S^1\times S^3$ can be regarded as TQFTs of a special type in this sense.
\vspace{0.2cm}

\noindent \emph{Superconformal indices}

In supersymmetric gauge theories, it is of most importance to understand the
spectrum of BPS states and the structures of moduli spaces. The Witten
index ${\rm Tr}\left(-1\right)^{F}e^{-\beta H}$ is a powerful
tool in counting the number of supersymmetric vacua since it is invariant
under the deformations of parameters of a theory. However, supersymmetric gauge theories have much richer structures so that the Witten index can only capture a little information. To extract more information, we need to harness symmetries of a theory. Fortunately, gauge theories, in general, flow to fixed points by renormalization group equations, ending up to become scale-invariant. In addition, it is believed that a scale-invariant theory of fields with spins less than one is conformally invariant \cite{Zamolodchikov:1986gt,Polchinski:1987dy,Dorigoni:2009ra,Antoniadis:2011gn}. As a result, the supersymmetry algebra is extended to the superconformal algebra. Thus, the study of SCFTs has a distinctive place in
the study of supersymmetric field theories as well as in the context of the AdS/CFT correspondence. Especially SCFTs on $\R\times S^3$ have been considerably investigated since the boundary of the five-dimensional anti-de Sitter space $AdS_5$ is $\R\times S^3$ and radial quantization of a SCFT on $\R^4$ is conformally mapped to the SCFT on $\R\times S^3$. An attempt to compute the partition function of $\N=4$ was first made in \cite{Sundborg:1999ue} and later extensively explored in \cite{Aharony:2003sx}.\footnote{
The study of supersymmetric gauge theory on $\R\times S^3$ traces back to the work by Diptiman Sen \cite{Sen:1985ph,Sen:1986zb,Sen:1989bg}. This should be appreciated since this work has not drawn  much attention although it is not directly related to the content here.}

In radial quantization, the Hilbert space of any unitary SCFT is decomposed into a direct sum over irreducible unitary representation of the superconformal algebra \cite{Dobrev:1985qv,Dobrev:1985vh,Dobrev:1985qz,Minwalla:1997ka,Dolan:2002zh}. Like  the highest weight representations, such representations are
classified by the BPS like conditions. These conditions are called the shorting and semi-shorting conditions, depending on how many supercharges annihilate states. The short and semi-short multiplets have the property that their energies are determined by the conserved charges that label the representation.

To count all the short and semi-short multiplets, the superconformal index, which is the generalization of the Witten index, was defined in \cite{Romelsberger:2005eg,Kinney:2005ej} by using the representations of the superconformal algebra. The index is constructed in such a way that it is independent of the continuous parameters of a theory. Hence, the evaluation of the index can be generally carried out  in a weakly-coupled limit \cite{Kinney:2005ej,Gadde:2009kb}. On the other hand, a large class of $\N=1$ SCFTs does not have weakly-coupled description. SCFTs of this kind naturally arise as IR fixed points of renormalization group flows, whose UV starting points are weakly-coupled theories.\footnote{We refer the reader to \cite{Strassler:2005qs} as a good exposition on this subject.} A prescription to evaluate the indices of such SCFTs was provided by R\"omelsberger \cite{Romelsberger:2005eg,Romelsberger:2007ec}. Yet apart from a number of checks for the duality correspondences \cite{Nakayama:2005mf,Nakayama:2006ur,Dolan:2008qi,Spiridonov:2008zr,Spiridonov:2009za,Spiridonov:2010hh,Gadde:2010en}, the reason why the prescription of R\"omelsberger works is not fully understood.

Nevertheless, there have been tremendous developments on the superconformal index, especially in the computational aspect. It was conjectured in \cite{Romelsberger:2007ec}
that the $\N=1$ indices for a Seiberg dual pair are identical. Invariance of the $\N = 1$ index under the Seiberg duality was systematically demonstrated in \cite{Dolan:2008qi,Spiridonov:2008zr,Spiridonov:2009za,Spiridonov:2010hh}. It appears
that superconformal indices are expressed in terms of elliptic
hypergeometric integrals. The identities between Seiberg dual pairs turn out to be equivalent to Weyl group symmetry transformations for higher order elliptic hypergeometric functions. 
Along this line, it was shown in \cite{Gadde:2009kb} that the index is invariant under the $S$-duality for the $\mathcal{N}=2$ SCFT with
$SU(2)$ gauge group and four flavors \cite{Seiberg:1994rs,Gaiotto:2009we}. Furthermore, using the inversion of the elliptic
hypergeometric integral transform, it was perturbatively tested in \cite{Gadde:2010te} that the
index for an interacting $E_6$ SCFT corresponds to the index of the $\N=2$ SCFT with $SU(3)$ gauge group and six flavors, providing a new evidence of the Argyres-Seiberg duality \cite{Argyres:2007cn}.

\vspace{0.2cm}

\ndt \emph{Functional integral interpretation and localization}

In this paper, we aim at providing the $\N=4$ superconformal index with geometric meaning. The $\N=4$  index is defined in a way that it counts the number of 1/16 BPS states in the $\N=4$ SCFT on $\R \times S^3$ that cannot combine into long representations under the deformation of any continuous parameter of the theory:
\be
{\cal I}={\rm Tr}(-1)^F \exp(-\beta \Delta) \ ,  \ \ \ \ \ \Delta\equiv2\{S,Q\}=H -2 J_3+ \tilde R_1+\tilde R_2+\tilde R_3
\label{introindex}
\ee
where $Q$ is one of supercharges and $S=Q^\dagger$.  Here $\{\tilde{R}_j\}_{ j=1,2,3}$ are an basis of the Cartan subalgebra of the $SU(4)_I$ $R$-symmetry. This can be regarded as the generalized Witten index. Like the Witten index, the $\N=4$ index can be interpreted as the Feymann functional integral with the Euclidean action by compactifying the time direction to $S^1$ with suitable twisted boundary conditions. Recalling that the $\N=4$ SCFT can be obtained by the dimensional reduction of the ten-dimensional $\N=1$ SYM, the form \eqref{introindex} of the $\N=4$ index tells us that the spatial manifold $S^3$ is rotated by the charge $J_3$ and the six-dimensional extra dimension $\IC^3$ is also rotated by the charge $\tilde R_j$ along the time direction, which is conventionally called a Scherk-Schwarz dimensional reduction. 

Our main purpose is to evaluate the functional integral with the Euclidean $\N=4$ SCFT action on this Scherk-Schwarz deformed background by using the localization technique. We shall show that the deformed action is $\delta_\e$-exact   where we choose $\e$ as the conformal Killing spinor which generates the fermionic symmetry $Q+S$. The functional integral reduces to the integral of one-loop determinants over the space of the critical points of the $\delta_\e$-exact term. Since there are no bosonic and fermionic zero modes due to the positive Ricci scalar curvature $R$, the functional integral is localized to zero modes of the gauge fields by integrating out all the other modes.

The main result is that the partition function for the $\N=4$ SCFT  on the Scherk-Schwarz deformed background with gauge group $G$ localizes to the following matrix integral:
\bea
{\cal I}(t,y,v,w) &=&
\int_{G} [dU]\,  \exp \left\{ \sum_{m=1}^\infty \frac 1m
f(t^m,y^m,v^m,w^m) \text{Tr}(U^\dag)^m \text{Tr}\,  U^m\right\}, \cr
f(t,y,v,w) &=& \frac{t^2(v+\frac 1w + \frac wv) - t^3 (y+\frac 1y)
- t^4 (w+\frac 1v+\frac vw) + 2 t^6}{(1-t^3y)(1-\frac{t^3}{y})}
\label{matrixintegral}
\eea 
where $f(t,y,v,w) $ is the character of the $PSU(1,2|3)$ subalgebra which commutes with $Q$ and $S$. This matches the result first obtained in \cite{Kinney:2005ej} by counting `letters' in the $\N=4$ SCFT on $\R\times S^3$. 
\vspace{0.2cm}

\ndt \emph{Plan of the paper}

The paper is organized as follows. In the section \ref{section2}, we review the rudiments of the ${\cal N}=4$ SCFT on $\R \times S^3$ with radial quantization. First we review the $\N=4$ index and explicitly write the Noether charges of the symmetries. Then we will re-derive a set of Bogomolnyi type equations for the bosonic 1/16 BPS configurations as found in \cite{Grant:2008sk}. The form of the Noether charge $\Delta$ suggests that this can be obtained as the energy of the system on an appropriate twisted background. In the section \ref{section3}, we provide the Feynman functional interpretation of the $\N=4$ index. The main thrust of this section is to find the action whose ``Hamiltonian'' is $\Delta$ by applying the methods in \cite{Nekrasov:2002qd,Pestun:2007rz}. This can be done by the dimensional reduction of the ten-dimensional $\N=1$ SYM on a Scherk-Schwarz deformed  background. The resulting action turns out to possess fermionic symmetries only generated by $Q$ and $S$. To implement the localization method, we provide the off-shell formulation of this system. In section \ref{section4}, we apply the localization method to evaluate the partition function on the Scherk-Schwarz deformed background. This section starts discussing the standard technique of localization. Then, we shall demonstrate that  the deformed action can be written as a $\delta_\e$-exact term. From the bosonic part of the $\delta_\e$-exact term, we show that the set of their critical points is the space of flat connections. It turns out that the space of flat connections on the Scherk-Schwartz deformed background can be regarded as the quotient space $T/W$ where $T$ is the maximal torus and $W$ is the Weyl group  of the gauge group $G$. We conclude this section by calculating the one-loop determinants around flat connections, which gives the desired matrix integral \eqref{matrixintegral}. The section \ref{section5} is devoted to conclusions and future directions and a number of technical points are detailed in the appendices.

\section{ ${\cal N}=4$ SCFT on $\R \times S^3$}\label{section2}
\subsection{$\N=4 $ Index}\label{SCI}
To begin with, we review the $\N=4$ superconformal index. We refer the reader to \cite{Kinney:2005ej} for more details as well as the Appendix \ref{appa} for the superconformal algebra. The $\mathcal{N}=4$ SCFT has the $PSU(2,2|4)$ space-time symmetry group which consist of the generators
\begin{equation}\left\{
\begin{array}{l}
H \\
J_a, \overline{J}_a \hspace{1.8cm}  a=1,2,3\\
P_\mu, Q_A^{\alpha},\overline{Q}^A_{\dot{\alpha}}, \ \ \ \ \  A=1,2,3,4 \\
K_\mu, S^A_\alpha,\overline{S}_A^{\dot{\alpha}} \ \ \ \ \ \ \ \alpha, \dot\alpha=\pm\\
\end{array}
\right.
\begin{array}{l}
{\rm dilations} \\
{\rm Lorentz\ rotations} \\
{\rm supertranslations} \\
{\rm special\ superconformal\ transformations} \ . \\
\end{array} 
\end{equation}
Just by convention, we call the supercharges $S^A_\alpha,\overline{S}_A^{\dot{\alpha}}$ superconformal charges. In radial quantization, these generators satisfy hermiticity properties. Especially, we have
\bea
\begin{array}{l}
S^A_\alpha=(Q^\alpha_A)^\dagger \\
\overline{Q}^{A\dot{\alpha}}=(Q_A^\alpha)^* \\ 
\end{array} \ \ \ \ \ \ 
\begin{array}{l}
\overline{S}_A^{\dot{\alpha}}=(\overline{Q}^A_{\dot{\alpha}})^\dagger\\
\overline{S}_{A\dot{\alpha}}=(S^A_\alpha)^* \\
\end{array}
\eea
where $*$ denotes complex conjugation and $\dagger$ denotes
Hermitian conjugation. 

Our main interest is to study 1/16 BPS states which are annihilated by the minimum number of supercharges, say, $Q\equiv Q^-_4$ and its hermitian conjugate $S\equiv S_-^4$. Before we discuss the $\N=4$ index, let us review the standard Hodge theory argument relating the $Q$-cohomology groups to the
ground states of $\Delta=2\{S,Q\}$. Because of $[Q,\Delta]=0$, the cohomology classes can be represented by eigenstates of $\Delta$. Consider a state $\xi$ such that $Q\xi=0$ with $\Delta \xi=\delta\xi$. If $\delta\neq 0$, then $ \xi=\frac 1\delta Q S\xi$ which means $\xi$ is $Q$-exact. Hence $Q$-cohomology groups always lie in the ground state of $\Delta$. Conversely, if $\Delta\xi=0$, then $0= \langle\xi |\Delta |\xi \rangle= 2|Q|\xi \rangle |^2+2|S|\xi \rangle |^2$ implies $Q|\xi \rangle=S|\xi \rangle =0$. If $\xi$ is $Q$-exact, more specifically $\xi=Q\zeta$, then  $\zeta$ is a zero eigenstate of $\Delta$ since $[Q,\Delta]=0$. This shows $\xi=Q\zeta=0$ by the argument just before. Therefore, we will consider the set of $1/16$ BPS states to be either all states that are annihilated by both $Q$ and $S$, or all states that are $Q$-closed but not $Q$-exact. 

Looking at the $\N=4$ superconformal algebra, $\Delta$ can be expressed by the sum of the quantum charges;
\begin{equation}
 \Delta\equiv2\{Q, S\}= H - 2J_3 + 2\sum_{k=1}^3 \frac k4R_k,
\label{susy}
\end{equation} 
where we denote the basis of the Cartan subalgebra of  the $SU(4)_I$ $R$-symmetry by $\{R_k\}_{k=1,2,3}$. $R_k$ may be thought of as the eigenvalues of the highest weight vectors under the diagonal generator $R_k$ whose $k^{th}$ diagonal entry is $1$, $(k+1)^{th}$ entry is $-1$,
and all the others are zero. For the later purpose, we define the matrix
\bea
\tilde T^A_{~B}\equiv2\sum_{k=1}^3 \frac k4R_k=\left(\begin{array}{cccc}
-\frac{1}{2}\\
 &- \frac{1}{2}\\
 &  &- \frac{1}{2}\\
 &  &  & \frac{3}{2}\end{array}\right) \ .
 \label{t}
\eea 
To count the 1/16 BPS states, the superconformal index is defined by 
\be
{\cal I} (t,y,v,w) =\Tr\left( (-1)^{\rm F}e^{-\beta\Delta}  t^{2(H+J_3)}
y^{2\overline{J}_3} v^{R_1} w^{R_2} \right)
\label{fugacity}
\ee
where fugacities $t,y,v,w$ are inserted to resolve degeneracies since $H+J_3, \overline J_3, R_2$ and $R_3$ commute with $Q$ and $S$. 
At zero coupling, the index can be evaluated by simply listing all basic fields or `letters' in the theory which have $\Delta=0$.
These are $\bar{\phi}_j$, $\chi^{\;\;\dot\a}_{\downarrow}$, $\lambda^{\;\;j}_{\uparrow-}$,
$(F^+)_{-}^{~+}$ (See \eqref{redefinition} and \eqref{SDFS} for notations) and derivatives $D_{+\dot{\alpha}}$ acting on them.
It turns out  from the superconformal algebra that these letters are $Q$-closed, but not $Q$-exact (See \eqref{BRSTQ}). The equation of motion for the $\N=1$ gaugino field
$
\partial_{+\dot{\a}}\chi^{\;\;\dot\a}_{\downarrow}=0
$
is only the equation of motion that can be constructed out of these letters.
Therefore, at zero coupling, any operator constructed out of the $\Delta=0$ letters, modulo this equation of motion,
will be $1/16$ BPS. The partition function over $1/16$ BPS states can be expressed as  the matrix integral \cite{Kinney:2005ej}
\bea \label{Ind} {\cal I}(t,y,v,w) =
\int_{G} [dU]\,  \exp \left\{ \sum_{m=1}^\infty \frac 1m
f(t^m,y^m,v^m,w^m) \text{Tr}(U^\dag)^m \text{Tr}\,  U^m\right\},\eea where $[dU]$ is
the $G=U(N)$ invariant Haar measure and $f(t,y,v,w)\text{Tr}\,  U^\dag \text{Tr}\,  U$ is so-called the
{\it single-particle states index}, or the {\it letter index} with
 \bea \label{spsi}
f(t,y,v,w) \ = \ \frac{t^2( w+\frac 1v+\frac vw)- t^3 (y+\frac 1y)
- t^4 ( v+\frac 1w + \frac wv) + 2 t^6}{(1-t^3y)(1-\frac{t^3}{y})}.\eea
It turns out that this single-particle partition function $f(t,y,v,w)$ is the character of the subalgebra $PSU(1,2|3)$ of the $PSU(2,2|4)$ symmetry. (See Eq.~(5.33) in \cite{Bianchi:2006ti}.) Therefore, this implies that the space of the 1/16 BPS states becomes an infinite dimensional representation space on which the subalgebra $PSU(1,2|3)$ acts.

It was shown in \cite{Kinney:2005ej} that the $\N=4$ index calculated in the free $\N=4$ SCFT with gauge group $U(N)$ using perturbation theory matches with the one computed in the strongly coupled $\N=4$ SCFT using the gravity description. Furthermore, generalizing this result, the whole list of the $\N=4$ superconformal indices with simple non-Abelian gauge groups are presented and  the invariance of superconformal index under exactly marginal deformations are shown in  \cite{Spiridonov:2010qv}.

\subsection{Action of ${\cal N}=4$ SCFT on $\R \times S^3$ and Noether Charges}\label{sec noether}
In this subsection, we review the basic properties of the ${\cal N}=4$ SCFT on $\R\times S^3$ \cite{Okuyama:2002zn,Ishiki:2006rt}. We refer the reader to the Appendix \ref{appa} and \ref{appb} for notations and conventions in detail. The action can be obtained by the dimensional reduction from the $\N=1$ SCFT in ten dimension $\R\times S^3 \times \IC^3$ where the ten-dimensional Lorentz group $SO(1,9)$ is decomposed to $ SO(1,3)\times SO(6)\subset SO(1,9) $:
\begin{eqnarray}
{\cal S}&=&\frac{1}{g_{YM}^2}
\int d^{10} x {\sqrt g}\; {\rm Tr}\left[\frac{1}{4} F_{MN}^2+\frac i2\bar{\lambda}\Gamma^MD_M\lambda+\frac{1}{12} RX_m^2\right]\cr
&=&\frac{1}{g_{YM}^2}\int d^4x {\sqrt g} \;  {\rm Tr}\Big[
\frac{1}{4} F_{\mu\nu}^2+\frac{1}{2}(D_{\mu}X_m)^2+  \frac{i}{2} \bar{\lambda}\Gamma^\mu D_\mu\lambda \cr
&& \hspace{4cm} +\frac{1}{2}\bar{\lambda}\Gamma^m[X_m,\lambda]+\frac{1}{4}[X_m,X_n]^2+\frac{1}{12}RX_m^2\Big]
\label{action}
\end{eqnarray}
where the ten-dimensional gauge fields $A_M$, $M=0,\cdots,9$  split the four-dimensional gauge field $A_\mu$, $\mu=0,\cdots,3$ and six scalars $X_m$, $m=1,\cdots,6$, and $\lambda$ is a ten-dimensional Majorana-Weyl spinor dimensionally reduced to the four dimension. Here $R=
\frac 6{r^2}$ is the Ricci scalar curvature of $S^3$.

Since the action is scaling invariant ${\cal S}[A_{\mu},X_m,\lambda,g_{\mu\nu}]= {\cal S}[A_{\mu},e^{-\alpha}X_m,e^{-3\alpha/2}\lambda,e^{2\alpha}g_{\mu\nu}]$, we can always choose the radius of the 3-sphere is equal to one, {\it i.e.}, $R=6$ for the Ricci scalar curvature. It is convenient to rewrite the action in the $SU(4)$ symmetric form:
\begin{eqnarray}
{\cal S}&=&\frac{1}{g_{YM}^2}\int d^4x{\sqrt g} \;  {\rm Tr}\Big[
\frac{1}{4}F_{\mu\nu}F^{\mu\nu}+\frac{1}{2}D_\mu X^{AB}D^\mu X_{AB}
+i\overline{{\lambda}_{\uparrow}}_{A}\gamma^{\mu}D_{\mu}{\lambda_{\uparrow}}^A+\frac{1}{2}X^{AB}X_{AB} \cr
&&+\overline{{\lambda}_{\uparrow}}_{ A}[X^{AB},\lambda_{\downarrow B}]+\overline{{\lambda}_{\downarrow}}^A[X_{AB},{\lambda_{\uparrow}}^B]
+\frac{1}{4}[X^{AB},X^{CD}][X_{AB},X_{CD}] \Big] , 
\label{actionSU(4)symmetric}
\end{eqnarray}
where $A,B=1, \cdots,4$, and the $\N=4$ gaugino $\lambda^A$ is transformed in the fundamental representation ${\bf 4}$ of $SU(4)_I$ $R$-symmetry, and the scalars $X^{AB}=-X^{BA}$ are in the antisymmetric tensor representation ${\bf 6}$ of $SU(4)_I$. In what follows, we shall use the $SU(4)$ symmetric form for the sake of the later arguments. The action is invariant under the superconformal transformations:
\begin{eqnarray}
&&\delta_{\epsilon}A_\mu=i(\overline{{\lambda}_{\uparrow }}_{A}\gamma_\mu{\epsilon_{\uparrow}}^A
-\overline{{\epsilon}_{\uparrow}}_{ A}\gamma_\mu{\lambda_{\uparrow}}^A), \cr
&&\delta_{\epsilon}X^{AB}=i(-\overline{{\epsilon}_{\downarrow}}^A{\lambda_{\uparrow}}^B
+\overline{{\epsilon}_{\downarrow}}^B{\lambda_{\uparrow}}^A+\epsilon^{ABCD}\overline{{\lambda}_{\uparrow}}_{ C}\epsilon_{\downarrow D}),
\cr
&&\delta_{\epsilon}{\lambda_{\uparrow}}^A=\frac{1}{2}F_{\mu\nu}\gamma^{\mu\nu}{\epsilon_{\uparrow}}^A
+2D_\mu X^{AB}\gamma^\mu \epsilon_{\downarrow B}+X^{AB}\gamma^\mu\nabla_\mu\epsilon_{\downarrow B}
+2i[X^{AC},X_{CB}]{\epsilon_{\uparrow}}^B, \cr
&&\delta_{\epsilon}\lambda_{\downarrow A}=\frac{1}{2}F_{\mu\nu}\gamma^{\mu\nu}\epsilon_{\downarrow A}
+2D_\mu X_{AB}\gamma^\mu{\epsilon_\uparrow}^B+X_{AB}\gamma^\mu\nabla_\mu{\epsilon_{\uparrow}}^B
+2i[X_{AC},X^{CB}]\epsilon_{\downarrow B} \ .
\label{susytrans}
\end{eqnarray}
where $\epsilon_{\updownarrow}=\frac12(1\pm\gamma_5)\epsilon$ and $\epsilon$ are conformal Killing spinors on $\R\times S^3$ satisfying 
\be
\partial_0\epsilon=\frac 12\gamma_0 \epsilon,\quad
\nabla_{i}\epsilon=\frac 12{\gamma}_{i}{\gamma}_{5} \epsilon \ .
\label{killing}
\ee
We note that the conformal Killing spinor equations \eqref{killing} can be obtained from the Killing spinor equations on $AdS_5$ restricted to the boundary $\R\times S^3$ \cite{Okuyama:2002zn}.
The supersymmetry is closed up to the equations of motion due to the on-shell formalism:
\be
[\delta_{\epsilon_1}, \delta_{\epsilon_2}]=\delta_{SO(2,4)}(\xi^\mu)+\delta_{SO(6)}(\Lambda^{mn})
+\delta_{{\rm gauge}}(v) +{\rm e.o.m.} 
\label{susyclosure}
\ee
where the parameters generating the corresponding symmetries are written by
\be
\xi^\mu=2i\bar{\epsilon_1}\Gamma^\mu\epsilon_2,\quad
\Lambda^{mn}=\frac i2\big(\bar{\epsilon_1}\Gamma^{mn}\Gamma^\mu \nabla_\mu\epsilon_2
-\bar{\epsilon_2}\Gamma^{mn}\Gamma^\mu \nabla_\mu\epsilon_1\big),\quad
v=-2i\bar{\epsilon_1}\Gamma^M\epsilon_2 A_M.
\label{parameter}
\ee
(For more explicit forms of the transformation by the square  $\delta_{\epsilon}^2$, see Eq. 2.7 and appendix C in \cite{Pestun:2007rz}.)

The stress-energy tensor $T_{\mu\nu}=(2/\sqrt{-g})(\delta(\sqrt{-g} {\cal L})/\delta g^{\mu\nu})$ is of form
\bea
T_{\mu\nu}&=&\frac{1}{g_{YM}^2}{\rm Tr} \left[\left(F_{ \mu\rho}F_{\nu}^{~\rho}-\frac 14 g_{\mu\nu}F_{\rho\sigma}F^{\rho\sigma}\right)+\left(D_\mu X^{AB}D_\nu X_{AB}-\frac12 g_{\mu\nu}D_\rho X^{AB}D^\rho X_{AB}\right) \right.\cr
&&+i(\overline{{\lambda}_{\uparrow}}_{ A}\gamma_{\mu}D_{\nu}{\lambda_{\uparrow}}^A+\overline{{\lambda}_{\uparrow}}_{ A}\gamma_{\nu}D_{\mu}{\lambda_{\uparrow}}^A-g_{\mu\nu}\overline{{\lambda}_{\uparrow}}_{A}\gamma^{\rho}D_{\rho}{\lambda_{\uparrow}}^A) -g_{\mu\nu}\left[\frac{1}{2}X^{AB}X_{AB}\right. \cr
&&+\left. \overline{{\lambda}_{\uparrow}}_{ A}[X^{AB},\lambda_{\downarrow B}]+\overline{{\lambda}_{\downarrow}}^A[X_{AB},{\lambda_{\uparrow}}^B]
+\frac{1}{4}[X^{AB},X^{CD}][X_{AB},X_{CD}]\right]
\label{se}
\eea
Then the Hamiltonian $H$ is given by 
\begin{eqnarray}
H &=& \int_{S^3} T_{00} \cr
&=&\frac{1}{ g_{{\rm YM}}^2}\int_{S^3}{\rm Tr}\left[
 \frac{1}{2}  F_{0j}^2+ \frac{1}{4}  F_{jk}^2+\frac{1}{2} |D_0 X^{AB}|^2+\frac{1}{2} |D_j X^{AB}|^2\right.\cr
&&+\  i\overline{{\lambda}_{\uparrow}}_{A}\gamma_{0}D_{0}{\lambda_{\uparrow}}^A+i\overline{{\lambda}_{\uparrow}}_{A}\gamma^{j}D_{j}{\lambda_{\uparrow}}^A+\frac 12 X_{AB}X^{AB} \cr
&&+\left.\overline{{\lambda}_{\uparrow}}_{ A}[X^{AB},\lambda_{\downarrow B}]+\overline{{\lambda}_{\downarrow}}^A[X_{AB},{\lambda_{\uparrow}}^B]+\frac{1}{4}[X_ {AB},X_{CD}][X^{AB},X^{CD}]
 \right].
 \label{hamiltonian}
\end{eqnarray}
Here the indices $ j,k=1,2,3$ run over a basis of the tangent space to $S^3$.

It is important to write down the Noether charges of the $PSU(2,2|4)$ symmetry explicitly. First, let us consider the conformal symmetry $SO(2,4)$. Let $M_{ab}$ be a conformal Killing vectors on $\R \times S^3$ satisfying
\begin{equation}
\nabla_{\mu}M_{\nu ab} +\nabla_\nu M_{\mu ab}=\frac{1}{2}(\nabla_\rho M^{\rho}_{ab})g_{\mu\nu} .
\end{equation}
On $\R \times S^3$, they obey the $SO(2,4)$ conformal algebra:
\begin{equation}
[M_{ab},M_{cd}]= i(\delta_{ad}M_{bc}-\delta_{bd}M_{ac}
-\delta_{ac}M_{bd}+ \delta_{bc}M_{ad}).
\end{equation}
where the indices $a,b,c,d$ run from $-2$ to $3$. 
The Noether current $j^\mu$ of a conformal Killing vector $M^\nu_{ ab}$ is given by $j^\mu_{ab}=T^{\mu\nu}M_{\nu ab}$.   To write the Noether currents of the $SU(2)_L\times SU(2)_R$ Killing vectors, it is often convenient to regard the 3-sphere $S^3$  as $SU(2)$ Lie group:
\begin{eqnarray}
SU(2) & = & \left\{ \left(\begin{array}{cc}
                          \a & \b \\
                          -\bar \b & \bar \a \\
                        \end{array}
                     \right) ; \a,\b \in \IC, |\a|^2+|\b|^2=1 
                     \right\}  \cr
                     &=& \left\{ g=e^{-i\phi \sigma^3/2}e^{-i\theta \sigma^2/2}e^{-i\psi \sigma^3/2}\right.\cr
                    &&\ \ \ \ \ =\left.\left(\begin{array}{cc}
                          \exp({-i\frac{\phi+\psi}{2}})\cos\theta & -\exp({-i\frac{\phi-\psi}{2}})\sin\theta \\
                          \exp({i\frac{\phi-\psi}{2}})\sin\theta  & \exp({i\frac{\phi+\psi}{2}})\cos\theta \\
                        \end{array}
                     \right);  0\le\phi, \psi \le 2\pi, \ 0\le \theta\le \pi
                     \right\} \ ,\cr 
                     &&\label{sphere}
                     \end{eqnarray}
            where we parametrize an element $g$ by the Euler angles $(\phi, \theta, \psi)$. Then the generators $J$ and $\overline J$ in \eqref{angmom} of the $SU(2)_L \times SU(2)_R$ symmetry are identified the left and right invariant vector fields on $SU(2)\cong S^3$ respectively. Under the isomorphism between the Lie algebra $\mathfrak{su(2)}\cong T_eSU(2)$ ($e$ is the identity element) and the left (right) invariant vector fields on $SU(2)\cong S^3$, we choose the Pauli matrices $\sigma^j$, $j=1,2,3$, (See \eqref{pauli}) as an orthonormal basis of the left (right) invariant vector fields where the metric is provided by the Cartan-Killing form.\footnote{We normalize the Cartan-Killing form as a symmetric bilinear form $\langle \ , \ \rangle :\mathfrak{su(2)}\times\mathfrak{su(2)} \to \IC; (g_1,g_2) \mapsto \frac 12 \Tr( g_1g_2)$. } Then the dual basis $e_{L(R)}^1, e_{L(R)}^2, e_{L(R)}^3$ of a left (right) invariant 1-form $\omega_{L(R)}$, so-called the left (right) invariant Maurer-Cartan forms, can be obtained by simple calculation
\bea
\omega_L=g^{-1}dg=i\sum_{j=1}^3 e_{L}^j \sigma^j , \ \ \ \  \omega_R=(dg)g^{-1}=i\sum_{j=1}^3 e_{R}^j \sigma^j 
\eea
where an element $g$ is as in \eqref{sphere} , and the dual orthonormal bases are written in terms of the coordinates $\theta, \phi, \psi$
\begin{equation}\left\{
\begin{array}{l}
e_L^1=\frac 12(\sin\psi d\theta-\cos\psi\sin\theta d\phi) \\
e_L^2=\frac 12(\cos\psi d\theta+\sin\psi\sin\theta d\phi) \\
e_L^3=\frac 12(d\psi+\cos\theta d\phi) \\
\end{array}
\right.
\left\{\begin{array}{l}
e_R^1=\frac 12(-\sin\phi d\theta+\cos\phi\sin\theta d\psi) \\
e_R^2=\frac 12(\cos\phi d\theta+\sin\phi\sin\theta d\psi) \\
e_R^3=\frac 12(d\phi+\cos\theta d\psi) \ . \\
\end{array}\right. \label{MC}
\end{equation}
They satisfy the Maurer-Cartan equations
\bea
de_L^j=\epsilon_{jkl}e_L^k\wedge e_L^l, \ \ \ de_R^j=-\epsilon_{jkl}e_R^k\wedge e_R^l \ .
\eea
In what follows, we choose  $(\partial/\partial x^0,2J_1,2J_2,2J_3)$ as an orthonormal basis of $\R \times S^3$ and focus only on the left invariant part. With the Euler angles $(\phi, \theta, \psi)$, the metric on $S^3$ is expressed as
\be
ds^2=\sum_{j=1}^3e_L^je_L^j=\frac 14\left[
d\theta^2+\sin^2\theta d\psi^2 +(d\phi+\cos\theta d\psi)^2\right] \ .
\label{metric2}
\ee
In addition, the left invariant vector fields $J_j,\  j=1,2,3$ are related to the coordinates $a=(\phi, \theta, \psi)$ via the dreibein $e_j^a$
\be
J_j =\frac 12 e_j^a \partial_a
\ee
where $e_j^a$ are the inverse metric of $(e_L)^{ j}_{ a} $. This identity gives us the explicit forms of the left invariant vector fields $J_j,\ j=1,2,3$ in terms of the Euler angles $(\phi, \theta, \psi)$
\bea
\left\{\begin{array}{l}
J_1=\sin\psi\partial_{\theta}+\cot\theta\cos\psi\partial_{\psi}-\frac{\cos\psi}{\sin\theta}\partial_{\phi} \cr
J_2=\cos\psi\partial_{\theta}-\cot\theta\sin\psi\partial_{\psi}+\frac{\sin\psi}{\sin\theta}\partial_{\phi} \cr
J_3=\partial_{\psi} \ . \cr
\end{array}\right. \label{leftinv}  
\eea
Then the Noether charge $J_3$ takes the form
\bea
J_3&=&  \int_{S^3} \frac12 T_{03} \cr &=&\frac{1}{g_{YM}^2}\int_{S^3} {\rm Tr} \left[\frac 12\left(F_{0\rho}F^{~\rho}_3+D_0 X^{AB}D_3 X_{AB}\right) +\frac  i2(\overline{{\lambda}_{\uparrow}}_{A}\gamma_{0}D_{3}{\lambda_{\uparrow}}^A+\overline{{\lambda}_{\uparrow}}_{ A}\gamma_{3}D_{0}{\lambda_{\uparrow}}^A) \right] \ .  \cr
&&\hspace{3cm}
\label{jcharge}
\eea
Here the coefficient in front of $T_{03}$ is determined by the norm $\|J_3\|=\frac 12$ which can be easily seen from the metric \eqref{metric2} and \eqref{leftinv}. The action is also invariant under the $SU(4)_I$ $R$-symmetry
\begin{equation}
\delta {\lambda_{\uparrow}}^A=iT^A_{~B}{\lambda_{\uparrow}}^B,\quad
\delta\overline{\lambda_\downarrow}_{ A}=-i\overline{\lambda_\downarrow}_{ B}T^B_{~A},\quad
\delta X^{AB}=iT^{A}_{~C}X^{CB}+iT^B_{~C}X^{AC},
\end{equation}
where $T^A_{~B}$ is a hermitian traceless matrix. The charge of this symmetry is
\begin{equation}
R^A_{~B}=\frac{1}{ g_{{\rm YM}}^2}
\int_{S^3}{\rm Tr}\Big(-i2X^{AC}D_0X_{CB}- \overline{{\lambda}_{\uparrow}}_{B}\gamma_0{\lambda_{\uparrow}}^A\Big).
\label{rcharge}
\end{equation}
Using \eqref{hamiltonian}, \eqref{jcharge} and \eqref{rcharge}, $\Delta=2\{S,Q\}=H-2J_3+ 2\sum_{k=1}^3 \frac k4R_k$ can be written as
\bea
\Delta&=&\frac{1}{ g_{{\rm YM}}^2}\int_{S^3}{\rm Tr}\left[
 \frac{1}{2} ( F_{0j}-F_{3j})^2+ \frac 12 F_{12}^2+\frac 12 |D_1 X^{AB}|^2+\frac 12 |D_2 X^{AB}|^2\right. \cr
 &&+2 |(D_0 -D_3+i)X^{j4})|^2- 2i(X^{j4}D_3X_{j4}-X_{j4}D_3X^{j4}  )\cr
&&+i\overline{{\lambda}_{\uparrow}}_{A}\gamma_{0}\{D_{0}-D_3+i\tilde T)\}{\lambda_{\uparrow}}^A+i\bar{\lambda}_{\uparrow A}\gamma^{j}D_{j}{\lambda_{\uparrow}}^A - i\overline{{\lambda}_{\uparrow}}_{A}\gamma_{3}D_{0}{\lambda_{\uparrow}}^A\cr
&&-\left. \overline{{\lambda}_{\uparrow}}_{ A}[X^{AB},\lambda_{\downarrow B}]+\overline{{\lambda}_{\downarrow}}^A[X_{AB},{\lambda_{\uparrow}}^B]+\frac{1}{4}[X^{AB},X^{CD}][X_ {AB},X_{CD}]
 \right].
 \label{noetherdelta}
\eea
where the indices $j=1,2,3$ run over the orthonormal basis of the left invariant vector fields as above and  $\tilde T=\tilde T^A_{~B}$ is defined in \eqref{t}. The bosonic part of the Noether charge corresponding to $\Delta$ can be expressed as a sum of squares (This was firstly derived in \cite{Grant:2008sk}. See section 4 and appendix C in \cite{Grant:2008sk}.)
\bea
\Delta_{{\rm Bosonic}}&=&\frac{1}{ g_{{\rm YM}}^2}\int_{S^3}{\rm Tr}\left[ \frac{1}{2} ( F_{0j}-F_{3j})^2+   \frac 12 \left(F_{12}+\frac 12[\phi^j,\bar \phi_j]\right)^2 \right.\cr
&& +\frac{1}{2} |(D_1+iD_2) \phi^j|^2+ \frac 12 |(D_0-D_3 +i)\phi^j|^2 +\frac14\sum_{j,k=1}^3\left| [\phi^j,\phi^k] \right|^2  \Big]. 
\eea
where the definitions of $\phi^j$ and $\bar \phi_j$ are given in \eqref{redefinition}.
Classical bosonic configurations with
$\Delta=0$ obey a set of first order Bogomolnyi equations obtained by
setting each of these squares to zero.
  The Bogomolnyi equations obtained in this way are
\begin{equation}
  F_{12}+\frac{1}{2}[\phi^j,\bar \phi_j]=0 \ ,\ \
  F_{0j}=F_{3j} \ \ (j=1,2,3)\ ,
  \label{moduli1}
\end{equation}
and
\begin{equation}
[\phi^j,\phi^k] =0\ ,\ \ (D_0-D_3 +i)\phi^j=0\ ,\ \
  (D_1+iD_2)\phi^j=0\ .
  \label{moduli2}
\end{equation}
This is a classical version of equation so that configurations
 satisfying the Bogomolnyi equations above preserve the supersymmetry generated by a single
supercharge and its Hermitian conjugate. The first equation in \eqref{moduli1} with the last one in 
\eqref{moduli2} is called the Hitchin equation. However, since the distribution spanned by $J_1, J_2$ is not involutive, it is not completely integrable on $S^3$ from the Frobenius theorem. Therefore the author does not know if there is a relation between the two-dimensional field theory and the $\N=4$ SCFT on $\R\times S^3$. 
Apart from the above set of the BPS equations, we should also impose the
Gauss law constraint to ensure the configuration solves all the equations of
motion:
\begin{equation}
  D^\mu F_{\mu 0}+\frac{i}{2}\left([\phi^j,D_0\bar\phi_j]+
 [ \bar\phi_j,D_0 \phi^j]\right)=0 \ .
\end{equation}

\section{Functional Integral Interpretation of $\N=4$ Index} \label{section3}
\subsection{Scherk-Schwarz Deformed Action}
In the last subsection \eqref{noetherdelta}, we can see that the time derivative $D_0$ is shifted to $D_0-D_3+i\tilde T$.  Heuristically, this implies that $S^3$ and the extra dimension $\IC^3$ are twisted along the time direction. Hence, in this section, we shall pursue the ${\cal N}=4$ index along this line of thought. 

Let us remind the meaning of the Witten index. The Witten index has Feynman functional integral interpretation
\be
\Tr(-1)^F e^{-\beta H}=\int_{\rm PBC} {\cal D} \Phi {\cal D} \Psi \exp[-{\cal S}_E(\Phi,\Psi)] \ ,
\ee
where the functional integral is taken over all the field configurations satisfying periodic
boundary conditions (PBC) along the compactified time direction $S^1$ with period $\beta$, and ${\cal S}_E$ is the Euclidean action of a theory.

Generalizing the Witten index, in \cite{Nekrasov:2002qd}, Nekrasov considered the equivariant index in the five-dimensional $\N=1$ SYM which is schematically written as
\be
\Tr (-1)^F e^{-\beta H} e^{\beta \Omega_{\mu\nu}J^{\mu\nu}}e^{\beta a^j R^j} \ ,
\label{equivindex}
\ee
where $H$ is the Hamiltonian,  $J^{\mu\nu}$ are generators of the $SO(4)$ rotation group and $R^j$ are generators of the Cartan subalgebra of the $R$-symmetry. By the functional integral interpretation mentioned above, this can be understood as the partition function on the five-dimensional manifold which is compactified on a circle with its circumference $\beta$ with twisted boundary condition $(t,x)\sim (t+\beta,\exp(i \beta\Omega_{\mu\nu}J^{\mu\nu})x)$ for $t\in S^1, x\in \R^4$. Here the operators $J^{\mu\nu}$  and $R^i$ preserve some of the supercharges of the theory which turns out to be topological charges. In the weakly coupled limit $\beta\to \infty$, the theory reduces to the supersymmetric quantum mechanics on the moduli space of instantons. The equivariant index \eqref{equivindex} is ended up with the integrals over the instanton collective coordinates which can be evaluated by the equivariant  localization. The resulting quantity may be conventionally called the instanton partition function $Z_{\rm inst}(\e_1,\e_2,a)$ where the parameters $(\e_1,\e_2)$ are the Cartan generators of the $U(1)^2$ rotation, and the parameter $a$ are those of a gauge group.  On the other hand, in the limit of $\beta\to 0$, the theory become the low energy effective theory of the $\N=2$ SYM in four dimensions. This consideration led to the conjecture that the low energy effective prepotential ${\cal F}(a)$ of the $\N=2$ SYM can be obtained from the instanton partition function: ${\cal F}(a)=-\lim_{\e_1,\e_2\to 0} Z_{\rm inst}(\e_1,\e_2,a)$ since the theory is topological and is independent of the coupling constant.\footnote{This conjecture was proven by the three groups independently \cite{Nekrasov:2003rj,Nakajima:2003pg,Braverman:2004cr}.}

\begin{figure}
\centering
    \includegraphics[width=6in]{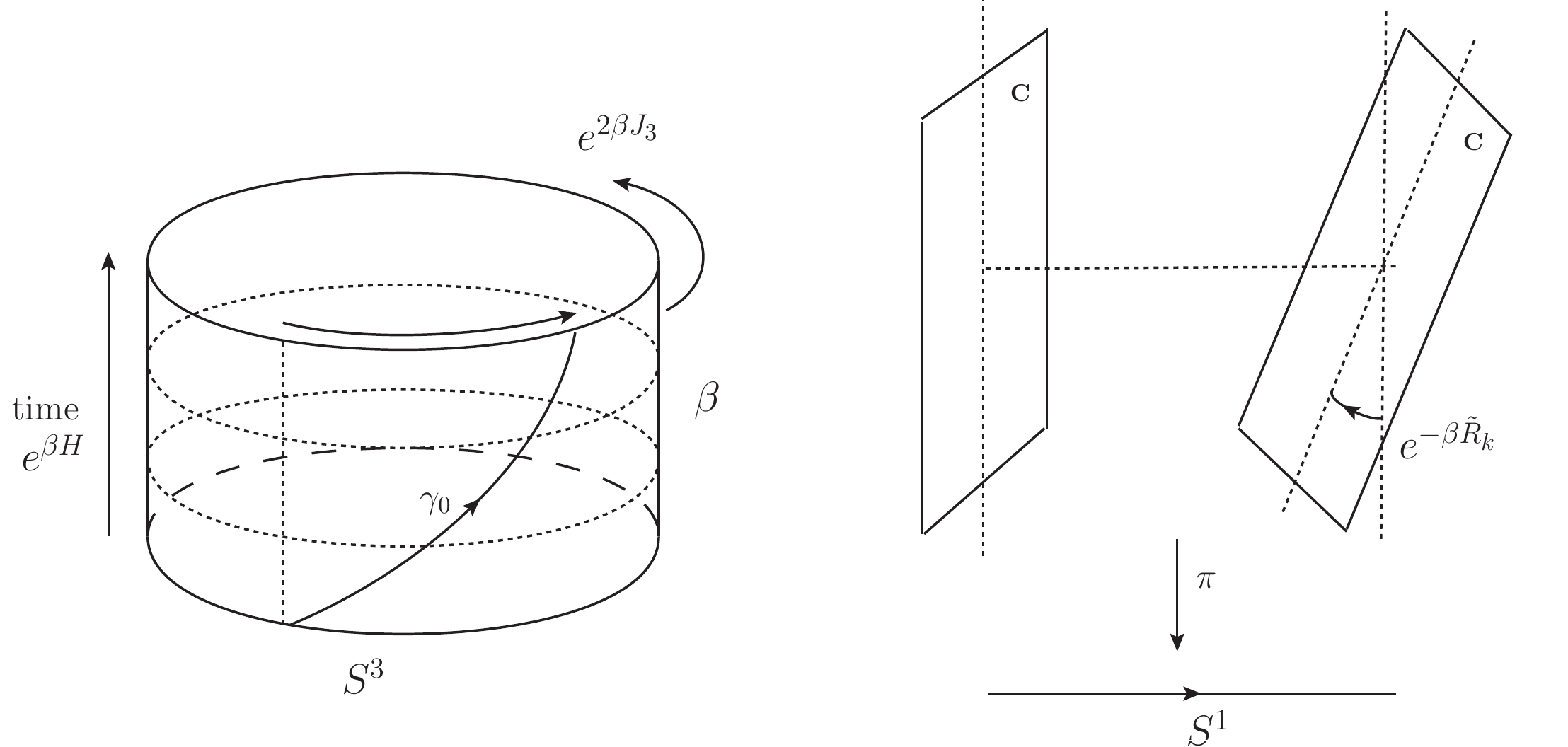}
  \caption{Schematic figures of the Scherk-Schwarz deformed background. Left: The 3-sphere at the top is identified with the one at the bottom by the rotation $e^{2\beta J_3}$. Here the time translation is vertical, which is common in physics literatures. The curve $\gamma_0$ depicts the integral curve of the vector field $\partial/\partial\tilde x^0$ which corresponds to the time direction of the space-time $M$. Right: the right 2-plane $\IC$ is identified with the right one by rotating $e^{-\beta \tilde R_k}$. Here the time direction $S^1$ can be seen as the base manifold of the fibre bundle with fibre a 2-plane $\IC$, which is common in mathematics literatures.}
  \label{fig omega} 
  \end{figure}

The results in \cite{Nekrasov:2002qd} are nice enough so that one may wonder if the
superconformal index can be interpreted in this way. This can indeed be done, and in a way that is closely related to the construction of the $\N=4$ SCFT from the ten-dimensional $\N=1$ SCFT by dimensional reduction. Recall that the ${\cal N}=4$ index is defined by
\begin{equation}
{\cal I}^{{\cal N}=4}={\rm Tr} (-1)^F e^{-\beta H} e^{2\beta J_3} e^{-\beta(\tilde R_1+\tilde R_2+\tilde R_3)} \ .
\label{n4index}
\end{equation}
Here we redefine, by $\{\tilde R_k\}_{k=1,2,3}$, the basis of the Cartan subalgebra of the $SU(4)_I$ $R$-symmetry.\footnote{$\tilde R_k$ may also be thought of as the eigenvalues of the highest weight vectors under the diagonal generator $\tilde R_k$ whose $k^{th}$ diagonal entry is $1/2$,  the forth entry is $-1/2$, and all the others are zero. The reason why we redefine the basis is to write the transition function \eqref{transfn} simply.} Since the $SO(6)\cong SU(4)$ $R$-symmetry comes from the Lorentz group $SO(1,9)$ in ten dimension, the part $e^{-\beta(\tilde R_1+\tilde R_2+\tilde R_3)} $ in the $\N=4$ index \eqref{n4index} rotates the extra dimensions $\IC^3$. This illustrates the fact that the $\N=4$ index \eqref{n4index} is nothing but an equivariant index of the ten-dimensional $\N=1$ SCFT. Thus, following \cite{Nekrasov:2002qd}, we shall interpret it as the partition function of the ${\cal N}=1$ SCFT on the fibre bundle $N$, more precisely $\xi=(N,\pi,S^1,S^3\times \IC^3)$, such that 
\begin{displaymath}
\xymatrix{ S^3\times\IC^3 \ar[r] & N \ar[d]_{\pi} \\
  & S^1  }
\end{displaymath}
where the twisted boundary condition, or the transition function, is given by 
\be
(x^0, \overrightarrow x, z^1, z^2, z^3) \sim (x^0+\beta, e^{2\beta J_3}\overrightarrow x,  e^{-\beta \tilde R_1}z^1, e^{-\beta \tilde R_2}z^2, e^{-\beta \tilde R_3}z^3)
\label{transfn}
\ee
(See Figure \ref{fig omega}).  Here we denote the local coordinates of the (4,7), (5,8), (6,9)-planes\footnote{We decompose the extra dimension $\IC^3$ to $\IC\times\IC\times\IC$ as follows. \newline
$$
\begin{tabular}{c|cccccccccc}
 & 0 & 1 & 2 & 3 & 4 & 5 & 6 & 7 & 8 & 9\tabularnewline
\hline
$z^{1}$ &  &  &  &  & $\times$ &  &  & $\times$ &  & \tabularnewline
$z^{2}$ &  &  &  &  &  & $\times$ &  &  & $\times$ & \tabularnewline
$z^{3}$ &  &  &  &  &  &  & $\times$ &  &  & $\times$\tabularnewline
\end{tabular}
$$} by $z^1,z^2,z^3$ respectively as consistent with \eqref{sixdimension} and \eqref{b4}. 

Let $\xi_1=(M,\pi_1,S^1,S^3)$, or simply $M$ for short, denote the subbundle of $N$ with fibre $S^3$ which is actually the space-time in this setting (the left of Figure \ref{fig omega}), and let $\xi_2=(L,\pi_2,S^1,\IC^3)$, or simply $L$ for short, denote the subbundle of $N$ corresponding to the rank 3 (complex) vector bundle over $S^1$ (the right of Figure \ref{fig omega}). The projection map $\pi:N\to S^1$ can be decomposed as $\pi=\pi_1\times \pi_2$.  With the transition function \eqref{transfn}, the connection of the vector bundle $L$ takes the value on ${\mathfrak u(1)}^3$. Since the transition function is properly normalized and the connection on $L$ is, in this case, independent of the coordinate of the base $S^1$, we can just consider the connection as $i$.\footnote{Here we use the fact that the Lie algebra $\mathfrak u(1)$ is isomorphic to $\sqrt {-1}$ $\R$} Keeping in mind that the fields $X^{j4}=\frac12\phi^j$ can be regarded as holomorphic sections of the vector bundle $L$ \footnote{More precisely, the fields $\phi^j$ are sections of the vector bundle $\mathfrak{g}_P\otimes L$ where $\mathfrak{g}_P$ is the adjoint bundle associated to a principal $G$-bundle over $M$, and $L$ can be regarded as the vector bundle over $M$, {\it i.e.} $\phi^j\in \Gamma(\mathfrak{g}_P\otimes L)$.}, $D_0 \phi^j$ has to be changed to $(D_0+i) \phi^j=D_0 X^m+M_{mn}X^n$ where $M_{mn}$ is the $\mathfrak{u(1)}^3$ subalgebra of the $\mathfrak{so(6)}$ $R$-symmetry and $m,n$ run over $1,\cdots 6$. Likewise, the derivative along the time direction acting on the $\N=4$ gaugino, $D_0 \lambda$, should be replaced by $(D_0+\frac14 \Gamma^{mn}M_{mn})\lambda=D_0 \lambda^A+i\tilde T^A_{~B} \lambda^B$.   This procedure is often described that a Wilson loop in the $R$-symmetry group is turned on (See section 2 in \cite{Nekrasov:2003rj}.).  Or, analogously, this construction is called the Scherk-Schwarz reduction of the ten-dimensional $\N = 1$ SCFT. 

On this twisted background, the time direction is shifted as $\frac{\partial}{\partial \tilde x^0}=\frac{\partial}{\partial x^0}+2J_3$ (the left in Figure \ref{fig omega}). We redefine the coordinate as
\be
\left\{
\begin{array}{l}
\frac{\partial}{\partial \tilde x^0}= \frac{\partial}{\partial x^0}+ 2J_3 \\
\tilde 2J^1=2J_1 \\
\tilde 2J^2=2J_2 \\
\tilde 2J^3=2J_3 \\
\end{array} \ \ \ \ \
\right.
\left\{\begin{array}{l}
 d\tilde x^0= d x^0  \\
\tilde e^1=e_L^1 \\
\tilde e^2=e_L^2 \\
\tilde e^3=e_L^3-  d x^0 \ . \\
\end{array}\right. \label{coordtrans}
\ee
where we note again that the norm $\|J_j\|$, $j=1,2,3$, is equal to $1/2$. 
 Hence the space-time $M$ is equivalent to the manifold with the topology $S^1\times S^3$ whose metric is given by \footnote{This equivalence is essentially the same as in the case of a 2-torus. The 2-torus with the flat metric $ds^2=\frac 12dwd\bar w$ and the periodicity $w\sim w+2\pi(m+n\tau)$ is equivalent the one with the metric $ds^2=|d\sigma^1+\tau d\sigma^2|$ and the periodicity $(\sigma^1, \sigma^2)\sim (\sigma^1, \sigma^2)+2\pi(m,n)$ for $m,n\in \Z$ (section 5.1 in \cite{Polchinski:1998rq})}
\bea
ds^2&=&(d\tilde x^0)^2+(\tilde e^1)^2+(\tilde e^2)^2+(\tilde e^3)^2 \\
      &=&(d x^0)^2+( e^1)^2+( e^2)^2+( e^3+  d x^0)^2   \ .
\label{twistedmetric}
\eea

 There is one subtlety that must be noted here. We obtained the Noether charge of $\Delta$ in the Minkowski signature in the subsection \ref{sec noether}. However, the Minkowski signature would be very embarrassing in the Hamiltonian treatment on $M$ since the vector $\partial/\partial \tilde x^0$ is  light-like  in the Minkowski signature. This consequently gives arise that the Noether charge of $H_{\rm twisted}$ is ill-defined due to $g_{00}=0$ as easily seen from the form of the stress-energy tensor \eqref{se}. Since the interpretation of the index by the Feynman functional integral is considered in the Euclidean signature, the Hamiltonian formulation in the twisted background is supposed to be carried out in the Euclidean signature too. Performing the Wick rotation on both the time coordinate and the connections simultaneously, the time derivative in this Scherk-Schwarz deformation of the Euclidean signature are consequently summarized in
\bea
D_0 \phi^j &\to& \left(D_0-2iJ_3+1 \right)\phi^j \cr
D_0 \lambda^j &\to& \left(D_0-2i\nabla_3- \tfrac12 \right)\lambda^j \cr
D_0 \lambda^4 &\to& \left(D_0-2i\nabla_3+ \tfrac32 \right)\lambda^4 \ .
\label{replace}
\eea
where $\nabla_3=J_3+\frac12\gamma^{12}$ since the spin connections change under the Scherk-Schwarz deformation
\be
\tilde\omega^{12}=\tilde e^3+2d\tilde x^0, \ \ \tilde\omega^{23}=\tilde e^1, \ \ \tilde\omega^{31}=\tilde e^1~.
\ee
This embraces the fact that there is no $i$ in the exponents of the rotation operators $e^{2\beta J_3}, \ e^{-\beta R_j}$. In other words, the rotation angles depicted in Figure \ref{fig omega} are purely imaginary.

The other thing we should bear in mind is the $\epsilon$-derivative terms $\nabla_\mu \epsilon$  in the superconformal transformation \eqref{susytrans}. Naively thinking, the action of the Scherk-Schwarz deformed $\N = 4$ theory on $M$ can be obtained by simply replacing the time derivatives as in \eqref{replace} with the metric \eqref{twistedmetric} on $M$. However, one needs to be careful that  the action obtained in this way is  invariant under the fermionic symmetry $Q$ and $S$ due to the $\epsilon$-derivative terms $\nabla_\mu \epsilon$  in the superconformal transformation \eqref{susytrans}. To see that, it is necessary to write the transformations  by $Q$ and $S$ on the Scherk-Schwarz deformed background explicitly.

Let us therefore look at conformal Killing spinors. The explicit solutions of the conformal Killing spinor equations \eqref{kspinorseq} depend on the choice of a metric and a vielbein. From the anti-commutation relation of  $Q^\a_A$ and $S_\a^A$, the vector fields $\epsilon^s \gamma^\mu \epsilon^q$ should be proportional to the Killing vectors $H$ (Hamiltonian) and $J_i$ (the left-invariant vector fields) of $S^1 \times S^3$. Hence, it is natural to choose $(\partial/\partial x^0,2J_1,2J_2,2J_3)$ as an orthonormal basis of the tangent space $T_pM$ for each point $p\in M$. We can write the conformal Killing spinor equations in the Euclidean signature which is compatible with this choice of coordinates can be written as \cite{Ishiki:2006rt}:
\bea
\nabla_\mu\e=\pm \frac 12\gamma_\mu\gamma^0\gamma^5\e \ ,
\label{kspinorseq}
\eea
or
\bea
\nabla_\mu \epsilon_\uparrow=\pm \frac{1}{2}\gamma_\mu \gamma^0 \epsilon_\uparrow,\;\;\;\;\; \nabla_\mu \epsilon_{\downarrow}=\mp \frac{1}{2}\gamma_\mu \gamma^0 \epsilon_{\downarrow} \ .
\label{conformalKillingequation}
\eea
where we have $\frac12$ in the right hand side due to the Euclidean signature instead of $\frac i2$ for the Minkowski signature as in \cite{Ishiki:2006rt}.
With this choice of the vielbein,  the spin connections $\omega^{ij}_k$ on $S^3$ is found to be $\omega^{ij}_k=\epsilon_{ijk}$ from \eqref{MC}. It turns out that the sign $\pm$ in \eqref{kspinorseq} agrees with the sign of the spin connection $\pm \e_{ijk}$ for left and right invariant vector fields on $S^3$. Then, it is straightforward to see that the solutions corresponding to $Q^\a_A$ and $S_\a^A$  can be written as
\bea
\epsilon^q=e^{\frac 12 x^0} \left(\begin{array}{c} \epsilon_0^q  \\ 0 \end{array}\right) \ \ \ \ \ \ \ \
\epsilon^s=e^{-\frac 12 x^0} \gamma^0 \left(\begin{array}{c}  \epsilon_0^s \\ 0 \end{array}\right)
\label{spinor solution}
\eea
where $\epsilon_0^q,\ \epsilon_0^s$ are covariantly constant spinors.\footnote{Although there are solutions of \eqref{kspinorseq} for the conformal Killing spinors corresponding to $\overline Q^A_{\dot\a}, \overline S_A^{\dot\a}$, they will be very tedious forms with this choice of the orthonormal frame. Since we are interested only in $Q$ and $S$, we do not obtain those solutions explicitly.} Unlike the Minkowski signature, the conformal Killing spinors \eqref{spinor solution} are not well-defined along the temporal circle $S^1$ although we need to impose the periodic boundary condition on them. This stems from the fact that $S^1\times S^3$ cannot preserve supersymmetry unless the theory is twisted according to \eqref{transfn}. However, for the time being, let us postpone this problem of the boundary condition on the temporal circle $S^1$.

To understand the meaning of the solution \eqref{spinor solution} more clearly, let us recall the spin representation of $Spin(10)$ in ten dimensions \cite{Pestun:2007rz}. The spinor space $\S_{32}$ in ten dimensions is of thirty-two dimensions which decomposes to the irreducible spin representations $\S_{16}^+,\ \S_{16}^-$ of $Spin(10)$ as $\S_{32}= \S_{16}^+ \oplus  \S_{16}^-$  ($\bf{32}_{\rm Dirac}=\bf{16}\oplus\bf{16^\prime}$). 
The gamma matrices $\Gamma^M$ in ten dimensions map a chiral spinor to the opposite chirality $\Gamma^M:\S_{16}^\pm \to \S_{16}^\mp$. The $\N=4$ gaugino $\lambda$ in \eqref{action} and the conformal Killing spinors $\e$ (not $\tilde \e$) satisfying \eqref{kspinorseq} take their value in $\S_{16}^+ $.  In the solutions \eqref{spinor solution}, the covariantly constant spinors  $\epsilon_0^q$ and $ \epsilon_0^s$ lie in $\epsilon_0^q\in \S_{16}^+ $ and $\epsilon_0^s\in \S_{16}^- $ which correspond to the generators of $Q^\a_A$ and $S_\a^A$ respectively. In addition, the generators $\epsilon_0^{\bar q}$ and $\epsilon_0^{\bar s}$ for $\overline Q^A_{\dot\a}$ and $ \overline S_A^{\dot\a}$ are contained as $\epsilon_0^{\bar q}\in \S_{16}^+ $ and $\epsilon_0^{\bar s}\in \S_{16}^- $.  In total, there are $32=16+16$ generators of the superconformal symmetry as expected. With this fact in mind, the relation between the solution \eqref{spinor solution} and the generator $ (\epsilon_0)_\a^A Q^\a_A+(\epsilon_0)_{\a A} S^{\a A}$ becomes more transparent:
\bea
\epsilon=\left(\begin{array}{c} \epsilon_\a^A \\ \bar\epsilon^{\dot\a}_A \end{array}\right) =e^{\frac 12 x^0} \left(\begin{array}{c} (\epsilon_0)_\a^A \\ 0 \end{array}\right) +e^{-\frac 12 x^0}\gamma^0  \left( \begin{array}{c}(\epsilon_0)_{\a A} \\ 0 \end{array}\right) =\left(\begin{array}{c}e^{\frac 12 x^0}  (\epsilon_0)_\a^A \\ e^{-\frac 12 x^0} (\bar\sigma^0)^{\dot\a \a}(\epsilon_0)_{\a A}   \end{array}\right) \ .
\label{explicit kspinors}
\eea
where  a conformal Killing spinor $\e$ takes its value in $\S_{16}^+$.

Now, in trying to see how the superconformal transformations \eqref{susytrans} are modified by the Scherk-Schwarz deformation, we shall rewrite the superconformal transformation \eqref{susytrans}  in terms of two-component spinor indices  
\begin{eqnarray}
&&\delta_{\epsilon}A_{\a \dot\a}=-2i(\overline{{\lambda_\uparrow}}_{\dot\a A}\epsilon_\a^A
-\bar{\epsilon}_{\dot\a A}\lambda_{\uparrow\a}^{\;\; A}), \cr
&&\delta_{\epsilon}X^{AB}=i(-{\epsilon}^{\a A }\lambda_{\uparrow\a}^{\;\; B}
+{\epsilon}^{\a B}\lambda_{\uparrow\a}^{\;\; A}+\epsilon^{ABCD}{\overline{ \lambda_\uparrow}}_{ C\dot\a}\bar\epsilon^{\dot\a}_{ D}),
\cr
&&\delta_{\epsilon}\lambda_{\uparrow\a}^{\;\; A}=F^{+~\b}_{\ \ \a}\epsilon_\b^A
+2(D_{\a \dot\a} X^{AB} )\bar\epsilon^{\dot\a}_{B}+X^{AB}\nabla_{\a\dot\a}\bar\epsilon^{\dot\a}_B+2i[X^{AC},X_{CB}]\epsilon_\a^B, \cr
&&\delta_{\epsilon} {\lambda_\downarrow}^{\dot\a}_A=F^{-\dot\a}_{\ \ \  \dot\b}\bar\epsilon^{\dot\b}_{A}
+2(D^{\dot\a \a} X_{AB} )\epsilon_{\a}^{B}+X_{AB}\nabla^{\dot\a\a}\epsilon_\a^B
+2i[X_{AC},X^{CB}]\bar\epsilon^{\dot\a}_{B}, 
\label{susytrans2}
\end{eqnarray}
where $F^{+~\b}_{\ \ \a}\equiv F_{\mu\nu}(\sigma^{\mu\nu})_\a^{~\b}$ and $F^{-\dot\a}_{\ \ \ \dot\b}\equiv F_{\mu\nu}(\bar\sigma^{\mu\nu})^{\dot\a}_{~\dot\b}$ are the self-dual and the anti-self-dual part of the gauge field strength (See Appendix \ref{appd}). Looking at the transformations of the $\N=4$ gaugino in \eqref{susytrans2},  the middle two terms are changed 
\bea
&&\left\{
\begin{array}{l}
2(D_{\a \dot\a} X^{k4} )\bar\epsilon^{\dot\a}_{4}=2\left[(\sigma^0)_{\a\dot\a} (D_0-2iJ_3+1)+(\sigma^j)_{\a\dot\a} D_j\right]X^{k4}\bar\epsilon^{\dot\a}_{4}\\
2(D^{\dot\a \a} X_{k4} )\epsilon_{\a}^{4}=2\left[(\bar\sigma^0)^{\dot\a\a} (D_0-2iJ_3-1)+(\bar\sigma^j)^{\dot\a\a}D_j \right]X_{k4}\epsilon_\a^4
\label{covariant derivative}
\end{array}\right . \\
&&\left\{
\begin{array}{l}
X^{k4}\nabla_{\a\dot\a}\bar\epsilon^{\dot\a}_4= X^{k4}\left[(\sigma^0)_{\a\dot\a} (\nabla_0-2i\nabla_3-\frac32)+(\sigma^j)_{\a\dot\a} \nabla_j\right]\bar\epsilon^{\dot\a}_4  \\
X_{k4}\nabla^{\dot\a\a}\epsilon_\a^4=X_{k4}\left[(\bar\sigma^0)^{\dot\a\a} (\nabla_0-2i\nabla_3+\frac32)+(\bar\sigma^j)^{\dot\a\a}\nabla_j \right]\epsilon_\a^4
\end{array}\right .
\eea
where the conformal Killing spinors, $\epsilon_\a^4$ and $\bar\epsilon^{\dot\a}_4$, generate the transformations by $Q_4^\a$ and $S^4_\a$. This introduction of the connections is compatible with the Leibniz rule. For instance, we have the identities
\bea
\left\{
\begin{array}{l}
(D_0-2iJ_3+1)\delta_\epsilon X^{k4}= [(\nabla_0-2i\nabla_3+\frac32)\epsilon^4]{\lambda_{\uparrow}}^k+\epsilon^4[ (D_0-2i\nabla_3-\frac{1}{2}) {\lambda_{\uparrow}}^k ] \\
(D_0-2iJ_3-1)\delta_\epsilon X_{k4}=[(\nabla_0-2i\nabla_3-\frac32)\bar\epsilon_4]\overline{\lambda_\uparrow}_k+\bar\epsilon_4[(D_0-2i\nabla_3+\frac12)\overline{\lambda_\uparrow}_k ]
\end{array} \right . \ . 
\eea

  Let us explicitly
compute superconformal variation of the Scherk-Schwarz deformed $\N=4$ theory on $\R\times S^3$. It is easy to see the variation in terms of the ten-dimensional Lagrangian \eqref{action} by using the superconformal transformation \eqref{susytrans3} with connection appropriately introduced by  the Scherk-Schwarz deformation. The variation can be read off up to the total derivative terms 
\bea
&& \delta_{\e} \left(  \frac 1 4 F_{MN} F^{MN} + \frac i2 \bar\lambda \Gamma^{M} D_{M} \lambda + \frac 1 {2r^2}
  X_{m} X^{m}\right)\cr 
  &=& i D_{M} (\bar \lambda \Gamma_{N} \e) F^{MN} + i\bar \lambda
  \Gamma^{M} D_{M} \left( \frac 1 2 F_{PQ} \Gamma^{PQ} \e -\frac 12 X_{m} 
  \Gamma^{m}  \nslash\e\right) + \frac i{r^2} (\bar\lambda \Gamma^{m} \e) X_{m} \cr
  &=& -
  i (\bar \lambda \Gamma_{N} \e) D_{M} F^{MN} + \frac i2\bar\lambda D_{M} F_{PQ} \Gamma^{M}
  \Gamma^{PQ} \e + \frac i2\bar\lambda \Gamma^M\Gamma^{PQ} D_M \e \;F_{PQ}  - \frac i2 \bar\lambda
  \Gamma^{M}  \Gamma^{m} \nslash \e \;D_{M} X_{m} \cr
&&-\frac i{2}
  \bar\lambda \Gamma^{M}  \Gamma^{m} D_M\nslash \e \;  X_{m} +\frac i{r^2} (\bar\lambda \Gamma^{m} \e) X_{m} \cr
  &=& -
  i (\bar \lambda \Gamma_{N} \e) D_{M} F^{MN} + \frac i2\bar\lambda D_{M} F_{PQ} \Gamma^{M}
  \Gamma^{PQ} \e + \frac i2\bar\lambda \Gamma^{Mm}\nslash \e \;F_{Mm}  - \frac i2 \bar\lambda
  \Gamma^{M}  \Gamma^{m} \nslash \e \;D_{M} X_{m} \cr
&&+ \frac i{2}
  \bar\lambda \Gamma^{m} \nslash^2\e \;  X_{m} +\frac i{r^2} (\bar\lambda \Gamma^{m} \e) X_{m} \cr
 && \label{variation}\eea
It is easy to see that the third and forth term cancel each other. In addition, by using the identity
\begin{equation}
  \label{eq:GammaMPQ}
  \Gamma^M \Gamma^{PQ} = \frac 1 3 (\Gamma^{M} \Gamma^{PQ} + \Gamma^{P} \Gamma^{QM} + \Gamma^{M} \Gamma^{PQ})
  + 2 g^{M[P} \Gamma^{Q]}
\end{equation}
and the Bianchi identity, we see that the first term cancels the second. (See around Eq. 2.23 in \cite{Pestun:2007rz} more in detail.) In the absence of the Scherk-Schwarz deformation, the Killing spinors satisfy
\be
\nslash^2\e=-\frac 13 R\e \ .
\label{killingid}
\ee
This identity can be obtained from the conformal Killing spinor equation \eqref{ckse} with the Weitzenb\"ock formula. Hence, the conformal Killing spinors obey $\nslash^2\e=-\frac 2{r^2} \e$ \footnote{This identity can also be obtained from the conformal Killing spinor equation \eqref{kspinorseq} once the radius $r$ of the 3-sphere $S^3$ is restored.} on $S^1\times S^3$ which leads the last two terms in \eqref{variation} cancel. However, this is no longer true after twisting the background since we have
\bea
\nslash^2\e&=&\left[\gamma^0\left(\nabla_0-2i\nabla_3-\frac32 \gamma^5\right)+\gamma^j\nabla_j \right]\left[\gamma^0\left(\nabla_0-2i\nabla_3+\frac32 \gamma^5\right)+\gamma^j\nabla_j \right]\left(\begin{array}{c} \epsilon_\a^4 \\ \bar\epsilon^{\dot\a}_4 \end{array}\right)\cr&=&\left(\frac{11}{4r^2}-\frac{6i}{r^2}\gamma^3\gamma^0\right)\left(\begin{array}{c} \epsilon_\a^4 \\ \bar\epsilon^{\dot\a}_4 \end{array}\right) 
\eea
where the relative sign difference in front of $\frac32 \gamma^5$ comes from the fact that $\nslash\e$ has the opposite chirality to $\e$, namely $\nslash\e\in\S^-_{16}$ and $\e\in\S^+_{16}$.
Therefore, the deformed Lagrangian is not invariant under $Q$ and $S$ naively:
\bea
\delta_\e \left(  \frac 1 4 F_{MN} F^{MN} + \frac i2 \bar\lambda \Gamma^{M} D_{M} \lambda + \frac 1 {2r^2}
  X_{m} X^{m}\right)= \left[\frac {19i}{8r^2} (\bar\lambda \Gamma^{m} \e)+3 (\bar\lambda \Gamma^{m} \Gamma^3\Gamma^0\e)\right] X_{m}\cr
\eea
This is a natural consequence of the Scherk-Schwarz deformation. (See Eq. 2.24 in \cite{Pestun:2007rz})  

To make the deformed action invariant under $Q$ and $S$, we need to chose a different conformal Killing spinor which satisfies \eqref{killingid} in the Scherk-Schwarz deformed background. In other words, we must solve the conformal Killing spinor equation with the derivative replaced by \eqref{replace}
\bea
\left[\p_0-2i\nabla_3+\frac32\gamma^5\right]\e&=&\frac12\gamma^5\e\label{s1}\\
\nabla_j\e&=&\frac12 \gamma_j\gamma^0\gamma^5\e \label{s3}
\eea
It easily follows from the algebra of the $\gamma$-matrices that a constant spinor solves the equation \eqref{s3} of $S^3$ part. Then, the equation  \eqref{s1} of $S^1$ part can be simplified to
\be
\p_0\e=-\left(1-i\gamma^3\gamma^0\right)\gamma^5\e~.
\label{another s1}
\ee
Using the basis of the $\gamma$-matrices chosen in \eqref{gamma euclid}, one finds that the solutions are of the form
\bea
\e=\left(\begin{array}{r}e^{-2 x^0}c_1\\c_2\\ c_3 \\  e^{2 x^0}c_4\end{array}\right)
\eea
where $c_i, \ i=1,2,3,4$ are constants. Since the first and forth components are not well-defined along the temporal circle $S^1$, we cannot choose them as supersymmetric generators. In fact, this implies that the generators for the supercharges $Q_4^+$ and $S_+^4$ are projected out due to the projection operator $1-i\gamma^3\gamma^0$ in the right hand side of \eqref{another s1}. Hence, only the supersymmetric generators for $Q$ and $S$ are well-defined in this Scherk-Schwarz deformed background. With the analogy to \eqref{explicit kspinors}, we can write the conformal Killing spinors which generate $Q$ and $S$ as
\bea
\epsilon=\left(\begin{array}{c} \epsilon_-^4 \\ \bar\epsilon^{\dot+}_4 \end{array}\right) =\left(\begin{array}{c} (\epsilon_0)_-^4 \\ 0 \end{array}\right) +\gamma^0  \left( \begin{array}{c}(\epsilon_0)_{+ 4} \\ 0 \end{array}\right) =\left(\begin{array}{c}  (\epsilon_0)_-^4 \\ (\bar\sigma^0)^{\dot++}(\epsilon_0)_{+ 4}   \end{array}\right) \ .
\label{explicit kspinors2}
\eea

With this choice of conformal Killing spinors, the twisted action is
\bea
{\cal S}_{\rm twisted}
  &=&\frac{1}{g_{YM}^2}\int_M d^4x{\sqrt g} \;  {\rm Tr}\Big[
\frac{1}{4}F_{\mu\nu}F^{\mu\nu}+\frac{1}{2}D_\mu X^{AB}D^\mu X_{AB}
+i\overline{{\lambda}_{\uparrow}}_{A}\gamma^{\mu}D_{\mu}{\lambda_{\uparrow}}^A+\frac {1} {2r^2}X^{AB}X_{AB} \cr
&&+\overline{{\lambda}_{\uparrow}}_{ A}[X^{AB},\lambda_{\downarrow B}]+\overline{{\lambda}_{\downarrow}}^A[X_{AB},{\lambda_{\uparrow}}^B]
+\frac{1}{4}[X^{AB},X^{CD}][X_{AB},X_{CD}] \Big] 
\label{twistedaction}
\eea
where the metric is as in \eqref{twistedmetric} and the time derivatives are  $(D_0-2iJ_3+\frac 1r) X^{j4}$, $(D_0-2iJ_3-\frac 1r) X_{j4}$  and  $(D_0-2i\nabla_3) \lambda^A+\frac 1r\tilde T^A_{~B} \lambda^B$. (Although we temporarily restore the radius $r$ of the 3-sphere here, we consider the case of  $r=1$ again in what follows unless it is explicitly mentioned.) According to the coordinate transformation \eqref{coordtrans}, the Hamiltonian $H_{\rm twisted}$ of this action \eqref{twistedaction} background is expressed as $H_{\rm twisted}=H- 2J_3$. With the extra dimensions rotated by the $R$-symmetry, we can identify as $H_{\rm twisted}=H- 2J_3+\sum_{k=1}^3 \tilde R_k=\Delta$.\footnote{Instead, we can think that the coordinate transformation in ten dimensions.
\be
\frac{\partial}{\partial \tilde x^0}=\frac{\partial}{\partial x^0}- 2J_3+\sum_{k=1}^3\frac{\partial}{\partial \theta^k}
\ee
where $\frac{\partial}{\partial \theta^k}$, $k=1,2,3$ is the vector fields which generate the rotations of $\IC\times\IC\times\IC$.Then under this coordinate transformation, it is easy to see $H_{\rm twisted}=H- 2J_3+\sum_{k=1}^3 \tilde R_i=\Delta$.} This twist of the extra dimensions (right figure in Figure \ref{fig omega}) breaks all the fermionic symmetries except $Q_4^\a$ and $S_\a^4$. Among them, $Q_4^+$ and $S_+^4$ are dropped by rotating $S^3$ by $J_3$ along the time direction (left figure in Figure \ref{fig omega}). Hence, only $Q$ and $S$ are left as fermionic symmetries of the Scherk-Schwarz deformed action \eqref{twistedaction}  as desired.

All in all, we can identify the $\N =4$ index as the partition function on the Scherk-Schwarz deformed background:
\be
{\cal I} =\Tr (-1)^F e^{-\beta \Delta}= \int_{\rm PBC} {\cal D}\Phi {\cal D} \Psi\exp(-{\cal S}_{\rm twisted}[  
\Phi, \Psi])
\label{index}
\ee
where $\Phi, \Psi$  stands for all the bosonic and fermionic fields  in the $\N=4 $ SCFT, respectively, and the periodic boundary condition is imposed on all the fields along the $\tilde x^0$-direction. 

\subsection{Off-shell Formulation}
To demonstrate the localization of the action \eqref{twistedaction} at the level of the functional integral and not just at the level of a classical action, we need an off-shell formulation of the fermionic symmetry of the theory \cite{Pestun:2007rz,Dabholkar:2010uh}.

In fact, it is easy to find an off-shell formulation in this case. To see that, let us discuss some properties of the Scherk-Schwarz deformed action \eqref{twistedaction}. With this choice of the supercharge  $Q\equiv Q^{-}_{4}$ and its hermitian conjugate $S\equiv S_{-}^{4}$, an $SU(3)\times U(1)$ subgroup of the original $SU(4)_I$ symmetry becomes manifest  in such a way that the $R$-symmetry indices $A=1,2,3$ express the representation $\bf 3$ of the $SU(3)$ part. This decomposition of the $SU(4)_I$ $R$-symmetry can be understood in terms of the  $\N=1$ superspace formulation of the $\N=4$ SYM on $\R^4$. From the point of view of $\N=1$ superspace, the $\N=4$ theory contains one $\N=1$ vector  multiplet $V$ and three $\N=1$ chiral multiplets $\Phi^j, j=1,2,3$, so that the physical component fields of these superfields are listed as
\be
V: (A_\mu, \lambda^4_{\alpha},\bar\lambda_4^{\dot\alpha}), \ \  \Phi^j: (\phi^j, \lambda^j_{\alpha}), \ \ \Phi^{\dagger}_j :(\bar\phi_j,\bar\lambda_j^{\dot\a})
\ee
where we can see that the representations of $SU(4)_I$ decompose according to $\bf6 \to \bf3\oplus\bf\bar3$, $\bf4 \to \bf3\oplus\bf1$. Recall that the $\N=4$ action on $\R^4$ takes the following form by the $\N=1$ superspace:
\bea
{\cal S}&=& \frac{1}{16g^2_{\rm YM}}\left[\int \!d^4 xd^2 \theta\, \Tr (W^2) + \int\! d^4 x d^2
\bar\theta\, \Tr (\overline W^2)\right] +\frac{1}{g_{\rm YM}^2} \int \!d^4 xd^2 \theta d^2
\bar\theta\, \Tr(\Phi^{\dagger}_j e^V \Phi^j) \cr   &&+\frac{i\sqrt2}{g_{\rm YM}^2}
\int \!d^4xd^2\theta \, \Tr\left\{\Phi^1[\Phi^2 ,\Phi^3]\right\} +
\frac{i\sqrt2}{g_{\rm YM}^2}\int \!d^4 xd^2\bar\theta\,\Tr\left\{\Phi^{\dagger}_1 
[\Phi^{\dagger}_2,\Phi^{\dagger}_3]\right\},
\label{n=1superspace}
\eea
where $W_\alpha =-\frac{1}{4}\bar D^2e^{-V}D_\alpha e^V$. The connection to the $\N=1$ superspace formalism stems from the fact that both $Q$ and $S$ lie in the $\N=1$ subalgebra as pointed out in \cite{Witten:1994ev}. 

With reference to the $\N=1$ superspace formalism, we can easily find an off-shell formulation. The action is modified with the quadratic term of the auxiliary fields $K^A, \ A=1,\cdots, 4$ 
\bea
{\cal S}_{\rm twisted}&=&\frac{1}{g_{YM}^2}\int_M d^4x{\sqrt g} \;  {\rm Tr}\Big[\frac 1 4 F_{MN} F^{MN} + \frac i2 \bar\lambda \Gamma^{M} D_{M} \lambda + \frac {1} {12}R
  X_{m} X^{m}+\frac12K^AK_A \Big] \cr
  &=&\frac{1}{g_{YM}^2}\int_M d^4x{\sqrt g} \;  {\rm Tr}\Big[
\frac{1}{4}F_{\mu\nu}F^{\mu\nu}+\frac{1}{2}D_\mu X^{AB}D^\mu X_{AB}
+i\overline{{\lambda}_{\uparrow}}_{A}\gamma^{\mu}D_{\mu}{\lambda_{\uparrow}}^A+\frac {1} {2r^2}X^{AB}X_{AB} \cr
&&+\overline{{\lambda}_{\uparrow}}_{ A}[X^{AB},\lambda_{\downarrow B}]+\overline{{\lambda}_{\downarrow}}^A[X_{AB},{\lambda_{\uparrow}}^B]
+\frac{1}{4}[X^{AB},X^{CD}][X_{AB},X_{CD}]+\frac12K^AK_A\Big] \cr
&&
\label{twistedaction2}
\eea
with the superconformal transformations
\begin{eqnarray}
\begin{array}{l}
\delta_{\epsilon}A_{\mu}=i(\overline{{\lambda_\uparrow}}_{ 4}\bar\sigma_\mu\epsilon^4
-\bar{\epsilon}_{ 4}\bar\sigma_\mu\lambda_{\uparrow}^{\;\; 4}) \ , \cr
\delta_{\epsilon}\phi^j=2i\e^4{\lambda_\uparrow}^j \ , \cr
\delta_{\epsilon}\bar\phi_j=2i\overline{\lambda_\uparrow}_j\bar\e_4 \ , \cr
\delta_{\epsilon}\lambda_{\uparrow\a}^{\;\; 4}=F^{+~\b}_{\ \ \a}\epsilon_\b^4
-\frac i2[\phi^j,\bar\phi_j]\epsilon_\a^4+ K^4 \e^4_\a \ , \cr
\delta_{\epsilon} {\lambda_\downarrow}^{\dot\a}_4=F^{-\dot\a}_{\ \ \  \dot\b}\bar\epsilon^{\dot\b}_{4}+\frac i2[\phi^j,\bar\phi_j] \bar\epsilon^{\dot\a}_{4}+K_4 \bar\epsilon^{\dot\a}_{4} \ , \cr
\delta_{\e} \lambda_{\uparrow\a}^{\;\; j}=(D_{\a \dot\a} \phi^j )\bar\epsilon^{\dot\a}_{4}+\frac12 \phi^j\nabla_{\a\dot\a}\bar\epsilon^{\dot\a}_4-\frac i2\epsilon^{jkl}[\bar\phi_{k},\bar\phi_{l}] \e^4_\a+K^j \e^4_\a \ , \cr
\delta_\e {\lambda_\downarrow}^{\dot\a}_j =(D^{\dot\a \a} \bar\phi_j )\epsilon_{\a}^{4}+\frac12\bar\phi_j\nabla^{\dot\a\a}\epsilon_\a^4 -\frac i2\epsilon_{jkl}[\phi^{k},\phi^{l}]\bar\epsilon^{\dot\a}_{4}  +K_j\bar\epsilon^{\dot\a}_{4} \ ,\cr
\delta_\e K^j=-2i\bar\epsilon_{4\dot\a}D^{\dot\a\a}{{\lambda}_{\uparrow}}^j_\a+2[\overline{\lambda_\uparrow}_4\bar\e_4,\phi^j] +\e^{jkl}[\overline{\lambda_\uparrow}_k\bar\e_4,\bar\phi_l] \ , \cr
\delta_\e K_j=-2i\epsilon^{4\a}D_{\a\dot\a}{\overline{\lambda_\uparrow}}_j^{\dot\a} +2[\e^4{\lambda_\uparrow}^4,\bar\phi_j] +\e_{jkl}[\e^4{\lambda_\uparrow}^k,\phi^l]  \ , \cr
\delta_\e K^4=\delta_\e K_4=iD_\mu\overline{{\lambda}_{\uparrow}}^{4}\sigma^\mu\epsilon^{4}-i\bar{\epsilon}_{ 4}\bar\sigma^\mu D_\mu\lambda_{\uparrow}^{\;\; 4}-  [\e^4{\lambda_\uparrow}^j,\bar\phi_j]  -[\overline{\lambda_\uparrow}_{j}\bar\e_4,\phi^j]\ , 
\end{array}
\label{susytrans5}
\end{eqnarray}
where $K_A=(K^A)^\dagger$, $K^4=(K^4)^\dagger=K_4$ and $K^j, \ j=1,2,3$ transform as the representation $\bf 3$ under the $SU(3)$ subgroup of the $SU(4)_I$ $R$-symmetry.

This off-shell formulation can also be obtained by using the Berkovitz method \cite{Berkovits:1993zz} in the dimensional reduction of the ten-dimensional $\N=1$ SYM (See section 4 in \cite{Evans:1994cb}).

\section{Localization}\label{section4}
In this section, we aim to compute \eqref{index} with the action \eqref{twistedaction2} in the off-shell formulation exactly by applying the localization method in TQFT.
 
 To give an inevitably very brief explanation of the localization method in TQFT, let us consider an infinite dimensional supermanifold $\mathcal{M}$ with an integration measure $d\mu$. Let $\delta_\e$ be a fermionic vector field on this manifold that such that $\delta_\e^{2}$ is a certain bosonic vector field  ${\cal L}_{\phi}$ and the measure is invariant under $\delta_\e$, {\it i.e}, $div_\mu \delta_\e$. The second property implies $\int_X \delta_\e f=0$ for any functional $f$ on $\mathcal{M}$. We would like to evaluate a functional integral of a $\delta_\e$-invariant action $ {\cal S}$ with some $\delta_\e$-invariant functional ${\cal O}$
\begin{equation}
Z({\cal O}) = \int_{\mathcal{M}} d\mu  \, {\cal O} \, e^{{- {\cal S}}} .
\label{pf}
\end{equation}
Suppose that the action can be written as a $\delta_\e$-exact term,  
$ {\cal S}=t\delta_\e U~$,
where $U$ is a fermionic, ${\cal L}_{\phi}$-invariant function and $t$ can be considered as a coupling constant. The variation of $Z$ with
respect to $t$ is
\be
\frac d{dt}Z({\cal O})= \frac d{dt}\int_{\mathcal{M}} d\mu\ {\cal O} e^{ - t \delta_\e U } = -\int_{\mathcal{M}} d\mu \{\delta_\e,U\} {\cal O} e^{ -t \delta_\e U} =- \int_{\mathcal{M}} d\mu \{\delta_\e,
U{\cal O} e^{- t \delta_\e U}\} = 0. 
\ee
In the limit of $t\to \infty$, the subspace $\mathcal{M}_{\e} \subset \mathcal{M}$ obeying $\delta_\e U=0$ only contributes the integral since the other configurations are exponentially suppressed.
In this limit, the integration for directions transverse to $\mathcal{M}_{\e}$ can be implemented exactly in the saddle point evaluation. Hence the integral is localized over the subspace $\mathcal{M}_{\e}$
\begin{equation}
Z({\cal O})= \int_{\mathcal{M}_{\e}} d\mu_{\e} \, {\cal O} \,  \, ,
\end{equation}
with a measure $d\mu_{\e}$ induced on the subspace $\mathcal{M}_\e$ by the original measure with the one-loop determinant.

In the present situation, we take the field space of the $\N=4$ SCFT in the off-shell formulation as $\mathcal{M}$ and the action \eqref{twistedaction2} as ${\cal S}_{\rm twisted}$, and we will not consider  observable $\cal O$ for the present. Since we have considered 1/16 BPS states,  $Q+S$ is chosen as a fermionic vector field $\delta_\e$. The conformal Killing spinor which generates $Q+S$ can be explicitly written from \eqref{explicit kspinors2}:
\bea
\epsilon=\left(\begin{array}{c} \epsilon_{-}^{4} \\ \bar\epsilon^{\dot+}_{4} \end{array}\right) =\left(\begin{array}{c}(\epsilon_0)_{-}^{4} \\  (\bar\sigma^0)^{\dot+ +}(\epsilon_0)_{+4}   \end{array}\right) \ .
\label{explicit killing3}
\eea
Following \cite{Pestun:2007rz}, we will take the following functional as $U$ so that the bosonic part of $\delta_\e U$ is positive definite:
\be
U=\int_Md^4x\sqrt g \; \frac14\Tr[({\delta_\e \lambda})^\dagger\lambda]
\label{CKS}
\ee
where $\lambda$ is the $\N=4$ gaugino.

\subsection{$\delta_\e$-Exact Term and Critical Points} \label{criticalpt}
In this subsection, we will explicitly show that  $\delta_\e U$ becomes the deformed action up to the coupling constant  and will find the set $\mathcal{M}_{\e}$ of the critical points of $\delta_\e U$.

In the flat space $\R^4$, the holomorphic part $\int \!d^2 \theta\, \Tr (W^2) $ of the gauge kinetic energy  can be written as $Q^{-}$-exact form, since $Q^{-}$ acts as $\int d\theta^{-}$  up to a total derivative \cite{Witten:1994ev}. Note that $Q_\a$ is expressed on $\R^4$ in terms of the $\N=1$ superspace formalism as $Q_\a= \partial_\a-i(\sigma^\mu\bar\theta)_\a\partial_\mu$. For instance, one can express the holomorphic  part as
\bea
\frac 14 \int \!d^2 \theta\, \Tr (W^2) &=&\frac14 Q^{-}\Tr[(Q^{+}\chi^{\a} )\chi_{\a})]\cr
   &=& \Tr\left[ \frac14 |F^+|^2+\frac i2 \bar{\chi} \Dslash \chi-\frac14D^2\right]\ .
\label{gaugekinetic}
\eea
where $\chi_\a\equiv\lambda^4_\a$  is the $\N=1$ guagino as defined in \eqref{redefinition}. For the same reason, the anti-holomorphic part  $  \int\!  d^2
\bar\theta\, \Tr (\overline W^2)$ can be written as $\overline Q_{\dot+}$-exact. The first line in the right-hand side of \eqref{gaugekinetic} looks similar to \eqref{CKS}. The corresponding part in $\delta_\e U$ can be expressed 
\bea
\frac 14 \delta_\e \Big[ (\delta_\e \chi_\uparrow)^\dagger\chi_\uparrow\Big]&=&\frac14\delta_\e\left[  -F_{\mu\nu} \e^{-}( \sigma^{\mu\nu}\chi_\uparrow)_{-}  +\left(\frac i2[\phi^j,\bar\phi_j]+K_4\right)\e^- {\chi_\uparrow}_{-}\right] \label{gaugekinetic3} \\
&=&- \frac14F_{\mu\nu}  F_{ \gamma\delta} \e^{-}(\sigma^{\mu\nu}\sigma^{ \gamma\delta})_-^{~-} \e_{-} -\frac i2D_\mu\left( (\overline{\chi_\uparrow} \bar\sigma_\nu)^- \e_--\bar\e^{\dot+}(\bar\sigma_\nu \chi_{\uparrow})_{\dot+} \right) \e^{-}( \sigma^{\mu\nu}\chi_\uparrow)_{-} \cr
&&+\frac1{4}\left(\frac 14|[\phi^j,\bar\phi_j]|^2+K^4K_4\right) \e^{-} \e_{-}-\frac14 [\overline{\lambda_{\downarrow}}^{j-}\e_-,\bar\phi_j]\e^- {\chi_\uparrow}_{-} -\frac14[\phi^j,\overline{\lambda_\uparrow}_{j\dot+}\bar \e^{\dot+}]\e^- {\chi_\uparrow}_{-} \cr
&&+\frac 14\left(i(D_{\mu}\overline{{\chi}_{\uparrow}}\bar\sigma^\mu)^{-}\e_--i(D_{\mu}{{\chi}_{\uparrow}}(\sigma^\mu)_{\dot+} \bar\epsilon^{\dot+}- [\overline{\lambda_{\downarrow}}^{j-}\e_-,\bar\phi_j]  -[\overline{\lambda_\uparrow}_{j\dot+}\bar \e^{\dot+},\phi^j]\right)\e^- {\chi_\uparrow}_{-}\cr
&& 
\label{gaugekinetic2}
\eea
where $\e_+=\e_{+4}\equiv( \epsilon_{-}^{4} )^\dagger=(\epsilon_0)_{+4} $ and we omit the indices $A=4$ in the conformal Killing spinors $\e$ for brevity. Since $\bar\e^{\dot+}=-(\bar\sigma^0)^{\dot++}\e_+$ cancels with $\e_+$, the terms which contain $\bar\e^{\dot+}$ vanish. In this case, \eqref{gaugekinetic2} is very similar to the flat case \eqref{gaugekinetic} except the $\e$-derivative $\nabla_\mu \e$ contribution in the second term of  \eqref{gaugekinetic2}:
\bea
\frac 14 \delta_\e \Big[(\delta_\e \chi_\uparrow)^\dagger\chi_\uparrow \Big]&=&\frac14 |F^+|^2+\frac i2 \overline{\chi_{\uparrow}}\Dslash \chi_\uparrow+\frac1{4}\left(\frac 14|[\phi^j,\bar\phi_j]|^2+K^4K_4\right)+\frac12 [\overline{\lambda_{\downarrow}}^{j},\bar\phi_j]{\chi_\uparrow}  \cr
&&\hspace{5cm} -[\e \ {\rm derivative \  contribution}]
\eea
where  we normalize $\e^-\e_-=1$ and the $\e$-derivative contribution is given by
\bea
&&[\e \ {\rm derivative\  contribution}]\cr
&&\hspace{1cm}=\frac i2\left[(\overline{\chi_\uparrow} \bar\sigma_\nu) \left(\nabla_0-2i\nabla_3+\frac{3}2\right)\e_-\right][\e^{-}( \sigma^{0\nu}\chi_\uparrow)_{-} ]+\frac i2[(\overline{\chi_\uparrow} \bar\sigma_\nu) (\nabla_j\e)_-][\e^{-}( \sigma^{j\nu}\chi_\uparrow)_{-} ]\cr
&&\hspace{1cm}=-4i  \overline{\chi_\uparrow}\bar\sigma^0\chi_\uparrow \ .
\label{e}
\eea
 The similar computation can be applied to the anti-holomorphic part
\bea
\frac 14 \delta_\e \Big[ (\delta_\e \chi_\downarrow)^\dagger\chi_\downarrow\Big]&=&\frac14 |F^-|^2+\frac i2 \chi_\downarrow\Dslash \overline{\chi_{\downarrow}} +\frac1{4}\left(\frac 14|[\phi^j,\bar\phi_j]|^2+K^4K_4\right)+\frac12[\overline{\lambda_\uparrow}_{j},\phi^j] {\chi_\downarrow}  \cr
&&\hspace{5cm} -[\bar\e \ {\rm derivative \  contribution}]
\eea
where the $\bar\e$-derivative contribution is given by
\bea
&&[\bar\e \ {\rm derivative\  contribution}]\cr
&&\hspace{1cm}=\frac i2\left[(\overline{\chi_\downarrow} \sigma_\nu) \left(\nabla_0-2i\nabla_3-\frac{3}2\right)\bar\e^{\dot+}\right][\bar\e_{\dot+}(\bar\sigma^{0\nu}\chi_\downarrow )^{\dot+} ]+\frac i2\left[(\overline{\chi_\downarrow} \sigma_\nu) (\nabla_j\bar\e^{\dot+})\right][\bar\e_{\dot+}(\bar\sigma^{j\nu}\chi_\downarrow )^{\dot+} ]\cr
&&\hspace{1cm}=4i\overline{\chi_\downarrow} \sigma^0\chi_\downarrow \ .
\label{bare}
\eea
Using the fact that $\chi_\downarrow=C_4(\overline{\chi_\uparrow})^T$ and  $ \chi_\uparrow=C_4(\overline{\chi_\downarrow})^T$ (See Appendix \ref{appb} for notation), one can convince oneself that \eqref{bare} cancels with \eqref{e}. Hence, we can  summarize the vector multiplet part of $\delta_\e U$
\bea
\delta_\e U \Big|_{\rm vect}&=&\int_M d^4x\sqrt g\;\Tr\left[\frac14 |F_{\mu\nu}|^2+ i \overline{\chi_{\uparrow}}\Dslash \chi_\uparrow+\frac 18|[\phi^j,\bar\phi_j]|^2+\frac12K^4K_4\right.\cr
&&\hspace{4cm}\left.+\frac12 \overline{\lambda_{\downarrow}}^{j}[\bar\phi_j,{\chi_\uparrow}]+\frac12\overline{\lambda_\uparrow}_{j}[\phi^j, {\chi_\downarrow}  ]\right] \ .
\label{vectmulti}
\eea

It is straightforward to compute the part of the three chiral multiplets in $\delta_\e U$ while one need to take care of $\e$-derivative contributions:
{\small
\bea
\delta_\e U \Big|_{\rm chiral}&=&\int_M d^4x\sqrt g\;\Tr\left[\frac14\delta_\e\left[(\delta_\e{\lambda_\uparrow}^j)^\dagger{\lambda_\uparrow}^j\right]+\frac14\delta_\e\left[(\delta_\e{\lambda_\downarrow}^j)^\dagger{\lambda_\downarrow}^j\right]\right]\\
&=& \int_M d^4x\sqrt g\;\Tr\left[\frac14\delta_\e\left[\bar\e_{\dot+}D^{\dot+\a}\bar\phi_j{\lambda_\uparrow}^j_\a+\frac12\bar\phi_j(\nabla_\mu\bar\e\bar\sigma^\mu)^{ \a}{\lambda_\uparrow}^j_\a+\frac i2\epsilon_{jkl}[\phi^{k},\phi^{l}]\e^-{\lambda_\uparrow}^{j}_-+K_j \e^-{\lambda_\uparrow}^{j}_-\right]\right. \cr
&&\hspace{1.8cm}\left.+\frac14\delta_\e\left[\e^{-}D_{-\dot\a}\phi^j{\lambda_\downarrow}_j^{\dot\a}+\frac12\phi^j(\nabla_\mu\e\sigma^\mu)_{\dot\a}{\lambda_\downarrow}_j^{\dot\a}+\frac i2\epsilon^{jkl}[\bar\phi_{k},\bar\phi_{l}] \bar\e_{\dot+}{\lambda_\downarrow}_{j}^{\dot+}+K^j \bar\e_{\dot+}{\lambda_\downarrow}_{j}^{\dot+}\right]\right]\label{secondline}\cr
&&\\
&=&\int_M d^4x\sqrt g\;\Tr\Big[\frac12 D_\mu\phi^jD^\mu\bar\phi_j+\frac12\phi^j\bar\phi_j+i\overline{\lambda_\uparrow}_j\;\Dslash{\lambda_\uparrow}^j+ \frac 18\sum_{j,k}|[\phi^j,\phi^k]|^2+\frac12K^jK_j\cr
&&\hspace{1.8cm}-\frac12\epsilon_{jkl}\overline{\lambda_\downarrow}^{k}[\phi^{l},{\lambda_\uparrow}^{j}]-\frac12\epsilon^{jkl}\overline{\lambda_\uparrow}_{k}[\bar\phi_{l},{\lambda_\downarrow}_{j}]-\frac12 \overline{\chi_{\downarrow}}[\bar\phi_j,{\lambda_\uparrow}^j] -\frac12\overline{\chi_\uparrow}[\phi^j, {\lambda_\downarrow}_j]\Big]\cr
\label{chiralmulti}
&&
\eea}
where the first term in the first line of \eqref{secondline} contains an $\e$-derivative contribution such as $[\bar\e_{\dot+}(\bar\sigma^\mu)^{\dot+\a}{\lambda_\uparrow}^j_\a][ \overline{\lambda_\uparrow}_{j\dot+}(D_\mu\bar\e^{\dot+})]$ which again turns out to cancel with the other $\e$-derivative contribution in the first term in the second line of \eqref{secondline}, $[\e^{-}(\sigma^\mu)_{-\dot\a}{\lambda_\downarrow}_j^{\dot\a}][\overline{\lambda_\downarrow}^{j-}(D_\mu\e_-)]$. Therefore, we find the bosonic part of $\delta_\e U$ as a sum of squares
{\small\bea
\delta_\e U&=&\int_M d^4x\sqrt g\;\Tr\Big[\frac14 |F_{\mu\nu}|^2+\frac12 D_\mu\phi^jD^\mu\bar\phi_j+\frac12\phi^j\bar\phi_j+\frac 18|[\phi^j,\bar\phi_j]|^2+\frac18\sum_{j,k}|[\phi^j,\phi^k]|^2 +\frac12K^AK_A\cr
&-&\!\!\!\left.\frac12\epsilon_{jkl}\overline{\lambda_\downarrow}^{k}[\phi^{l},{\lambda_\uparrow}^{j}]\!-\!\frac12\epsilon^{jkl}\overline{\lambda_\uparrow}_{k}[\bar\phi_{l},{\lambda_\downarrow}_{j}]\!-\!\frac12 \overline{\chi_{\downarrow}}[\bar\phi_j,{\lambda_\uparrow}^j] \!-\!\frac12\overline{\chi_\uparrow}[\phi^j, {\lambda_\downarrow}_j]\!+\!\frac12 \overline{\lambda_{\downarrow}}^{j}[\bar\phi_j,{\chi_\uparrow}]\!+\!\frac12\overline{\lambda_\uparrow}_{j}[\phi^j, {\chi_\downarrow}  ]\right] \ . \cr
&&
\eea}
Thus the action itself can be written as a $\delta_\e$-exact form as expected.
\be
{\cal S}_{\rm twisted}=\frac1{g_{\rm YM}^2}\delta_\e U
\ee
This explains the reason why the $\N=4$ index is independent of the coupling constant. The set $\mathcal{M}_\e$ of the critical points of $\delta_\e U$ is the space of flat connections $F_{\mu\nu}$ with $\phi^j=0, \ K^A=0$. This result can be understood in the following way. There are no zero modes of the scalar fields $\phi^j$ since there are the curvature coupling terms in the Lagrangian. Besides, the Weitzenb\"ock formula $\nslash^2=\Delta+\frac R4$ tells us that there are no zero fermionic modes since the Ricci scalar curvature $R$ is positive. Hence, the solution we found illustrates the fact that we can integrate out all the fields in the functional integral except zero modes of the gauge fields which are, in fact, flat connections. This conclusion can be also obtained by using the superconformal transformation by $Q$ and $S$. (See Appendix \ref{appc})  To even make one step further, let us suppose that  we add the $\theta$-angle to the action ${\cal S}_{\rm twisted}$
\be
\frac{i\theta}{16\pi^2}\int_M d^4x\sqrt g \; \Tr F\wedge F~.
\label{theta}
\ee
Then we can regard $e^{\frac{i\theta}{16\pi^2}\int_M d^4x\sqrt g \; \Tr F\wedge F}$ as an observable $\cal O$ in \eqref{pf}. Since only the flat connections make contributions to the functional integral in the weak coupling limit of $g_{\rm YM}\to0$ and the term \eqref{theta} vanishes on the space $\mathcal{M}_\e$ of flat connections,  the $\N=4$ index turns out to be also independent of the $\theta$-angle.

\begin{figure}
\centering
    \includegraphics[width=2in]{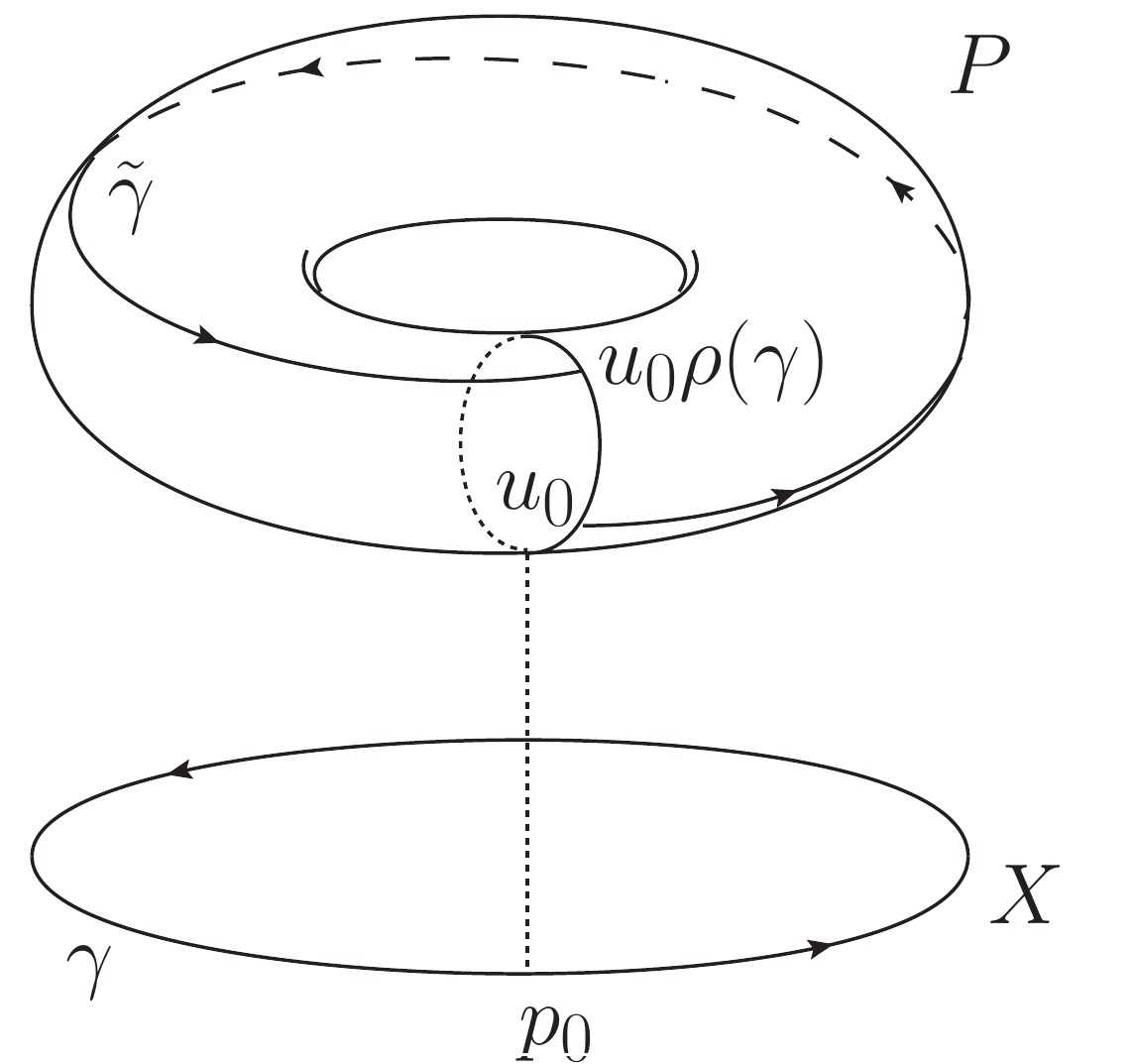}
  \caption{A schematic figure which explains the geometric meaning of a Wilson line.}
  \label{fig holonomy} 
  \end{figure}

Let us therefore make a few remarks about flat connections. The geometric meaning for a connection $A$ to be flat can be explained by using the theorem of Frobenius as follows. For each point $p\in P$ of the principal $G$-bundle $P$, we define 
\be
{\cal H}_u=\{v\in T_uP;A(v)=0\} .
\ee
${\cal H}$ is a distribution consisting of all horizontal vectors relative to $A$. Then the sufficient and necessary condition for a connection $A$ to be flat is that the distribution ${\cal H}$ to be completely integrable. This fact tells us that for any closed curve $\gamma$ which starts at $p_0\in X$ in the base manifold $X$ there is a unique lift $\tilde \gamma$ starting at $u_0\in \pi^{-1}(p_0)$ and lying in the integral manifold of $\cal H$ through $u_0$. The end point of $\tilde\gamma$ lies in the same fiber $\pi^{-1}(p_0)$ as $u_0$. Thus, there exists an element $g\in G$ such that the end point of $\tilde \gamma$ can be expressed as $u_0g$. (See Figure \ref{fig holonomy}) Then, it turns out that the element $g$  depends only on the homotopy class of the closed curve $\gamma$. Therefore, by setting $\rho(\gamma)=g$, we can define a map 
\be
\rho:\pi_1(X)\to G \ .
\ee
In fact, it is easy to show that this map is a homomorphism, which is called a {\it holonomy homomorphism}. Next, suppose that we choose a different point $u'_0\in \pi^{-1}(p_0)$ in the same fiber instead of $u_0$. Then there exists $h\in G$ such that $u'_0=u_0 h$.  Then, the resulting holonomy homomorphism $\rho'$ constructed as above is related to $\rho$ by conjugacy; $\rho'(\gamma)=h\rho (\gamma)h^{-1}$. Since the connection $A$ defines the horizontal direction on the fiber bundle, the holonomy homomorphism $\rho(\gamma)$ can be put as $P\exp \oint_\gamma A$ where $P$ indicates path-ordering. From the holonomy homomorphism, we obtain a Wilson loop operator $W_R(\gamma)=\Tr_R P\exp \oint_\gamma A$ by taking the trace of this element. Note that a Wilson loop is independent of the choice of a starting point on the fiber.

It is known that the structure of a flat bundle is completely defined by its holonomy. Namely, there is one-to-one correspondence between the space of the flat connections on $X$ and  the set of conjugacy classes of the homomorphism $\rho:\pi_1(X)\to G$.

Returning to the case at hand, the fundamental group $\pi_1(M)$ is homomorphic to $\Z$ represented by the time circle $S^1$ (the integral curve $\gamma_0$ in  Figure \ref{fig omega}). Hence, each flat connection corresponds to a holonomy group $\rho(\gamma_0)\in G$ up to conjugacy. Recall that, given a maximal torus $T$ in $G$, every element $g \in G$ is conjugate to an element in $T$, and the Weyl group $W$ acts on the maximal torus $T$ as an automorphism group. Therefore, the space of the flat connections on $M$ can be identified with $T/W$. In the case of the $U(N)$ gauge group, the maximal torus $T$ is isomorphic to an $N$-torus $\overbrace{S^1\times \cdots\times S^1}^N$, and the Weyl group $W$ is the symmetric group $\mathfrak{S}_N$ of degree $N$ whose action on $T$ is given by
\bea
W&:&T\to T\cr
&;&t={\rm diag}(t_1, \cdots,t_N)\mapsto w\cdot t:={\rm diag}(t_{w(1)}, \cdots,t_{w(N)})
\eea
where $t_i\in \IC, \ i=1,\cdots N$ with $|t_1|=\cdots=|t_N|=1$ and $w\in\mathfrak{S}_N$. The space of the $U(N)$ flat connections on $M$ is the quotient space $(S^1\times \cdots\times S^1)/\mathfrak{S}_N$.

Since the Scherk-Schwarz deformed action \eqref{twistedaction2} vanishes on the set $\mathcal{M}_\e$ of the critical points, the index can be exactly implemented by the one-loop evaluation of $\delta_\e U$ on the space of flat connection $T/W$:
\bea
{\cal I}^{\N=4}=\frac{1}{\# W}\int_T [dU] Z_{\rm 1-loop}
\label{localizedpf}
\eea
where $\# W$ is the order of the Weyl group $W$ (For the $U(N)$ gauge group, $\# \mathfrak{S}_N=n!$) and $[dU]$ is the Haar measure on the maximal torus $T$.

\subsection{One-Loop Evaluations}\label{4.3}
In this subsection, we compute the one-loop determinants coming from quadratic fluctuations of the fields about the flat connections on $M$. In the limit of $g_{\rm YM}\to0$, it is enough to keep only quadratic
terms in the bosonic fields $ \Phi=(A, \phi)$ and fermionic fields $\Psi=(\chi,\lambda)$. The quadratic terms are of the general form,
\be
{\cal L}=\int_M\sqrt g (\Phi\Delta_B\Phi+i\Psi D_F\Psi)
\ee
where $\Delta_B$ and $D_F$ are certain second and first order differential operators, respectively.  The Gaussian integral over $\Delta_B$ and $D_F$ gives
\be
Z_{\rm 1-loop}=\frac{\rm Pfaff \ D_F}{\sqrt{\det \Delta_B}}
\ee
where Pfaff denotes the Pfaffian of the real, skew-symmetric operator $D_F$. We will mainly follow the arguments made in the section 4 of \cite{Aharony:2003sx} and in the appendix B of \cite{Kim:2009wb} to demonstrate the one-loop evaluation explicitly.

In attempting to examine the saddle-point evaluation of the vector multiplet part \eqref{vectmulti}, we first fix the gauge. Following \cite{Aharony:2003sx}, we take the Coulomb gauge $\nabla_ jA^j=0$. The residual gauge symmetry is fixed by 
\be
\frac d{dt}\a(t)=0, \ \ \ \  \a\equiv\frac 1{\omega_3}\int_{S^3} A^0
\ee
where $\omega_3$ is the volume of $S^3$. For the residual gauge,
the Faddeev-Popov determinant is given by $ \prod_{\substack{k<l}}\left[2\sin
  \frac{\alpha_k\!-\!\alpha_l}{2}\right]^2$, which provides  the Haar measure $[dU]$ on the maximal torus of $U(N)$ in \eqref{localizedpf}:
\bea
\int_T [dU] \to \prod_{k=1}^N \int_{-\pi}^{\pi} d \alpha_k
\prod_{k<l} \sin^2\left({{\alpha_k - \alpha_l}\over 2}\right)
\eea
where $\a={\rm diag}(\a_1,\cdots,\a_N)$. With the Faddeev-Popov measure, the one-loop partition function can be written as 
\bea 
 Z_{\rm 1-loop}^{\rm vect}=\int {\cal D}A_0 {\cal D} A_j{\cal D}c{\cal D}\bar c\; \;\delta(\nabla_ jA^j)e^{-{\cal S}_0^{\rm vect}}
 \eea
Here, keeping only the quadratic terms of \eqref{vectmulti}, we denote the gauge-fixed action with the Faddeev-Popov ghosts $c, \ \bar c$  by ${\cal S}_0^{\rm vect}$
\bea
{\cal S}_0^{\rm vect}&=&\int_Md^4x\sqrt g \; \Tr\left[ -\frac12 A_j (\tilde{D}_0^2 +\nabla^2) A^j +
-\frac12 A_0 \nabla^2 A_0 + \frac i2\overline{\lambda_\uparrow} \nslash \lambda_\uparrow -{\bar c} \nabla^2c\right]
\eea
 where $\tilde{D}_0 A_j\equiv \p_0 A_j -i[\alpha, A_j]$.  
To go further, we decompose the gauge field into a pure divergent  and a divergenceless as $A_j= \p_j \varphi + B_j$ where $\p_j B^j=0$. Then the delta function constraint becomes $\delta(\nabla^2\varphi)$ and the integral over $\varphi$ yields $[\det'(\nabla^2) ]^{-1/2}$ where the
derivatives act on scalar functions on $S^3$ and the prime indicates that zero modes are not counted. The integral over $A_0$ yields the same factor. The integral over
the ghosts, on the other hand, evaluates to $\det'(\nabla^2)$. These three
factors cancel nicely, and we are left with
\bea
{\cal S}_0^{\rm vect}&=&\int_Md^4x\sqrt g \; \Tr\left[-\frac12B_j (\tilde{D}_0^2 +\nabla^2) B^j+ \frac i2\overline{\lambda_\uparrow} \nslash \lambda_\uparrow 
\right] \ .
\label{gaugeinv}
\eea

Before performing the Gaussian integral of \eqref{gaugeinv}, let us discuss about the insertion of the chemical potentials to the $\N=4$ index as in \ref{fugacity}.
We shall replace the fugacities $t,y,v,w$ in \eqref{fugacity} by chemical potentials $\tau, \gamma,\zeta_1,\zeta_2$:
\be
{\cal I}(\tau,\gamma,\zeta_1,\zeta_2)=\Tr (-1)^F e^{-\beta\Delta}e^{-2\tau(E+J_3)}e^{-2\gamma \overline J_3} e^{-\zeta_1 R_1}e^{-\zeta_2 R_2}
\ee
where $ t= e^{-\tau},y=e^{-\gamma},v=e^{-\zeta_1},w=e^{-\zeta_2}$.
Then the insertion of the chemical potentials induces additional twist of the background which can be addressed by
replacing all time derivatives in the action by 
\be
D_{0}\rightarrow\partial_{0}-\frac i{\beta\!+\!2\tau}[\a, \ ]+
  \frac{2(\tau\!-\!\beta)}{\beta+2\tau}(i\nabla_3)
  +\frac{2\gamma}{\beta\!+\!2\tau}(i\overline \nabla_3) +\frac{\frac12\beta\!+\!\zeta_1}{\beta\!+\!2\tau} R_1
  +\frac{\beta\!+\!\zeta_2}{\beta+2\tau}R_2+\frac{\frac32\beta}{\beta\!+\!
  2\tau}R_3
  \label{time chemical}
\ee
Since $R_1$, $R_2$ and $\overline \nabla_3$ act trivially on the conformal Killing spinor $\left(\begin{array}{c} \epsilon_\a^4 \\ \bar\epsilon^{\dot\a}_4 \end{array}\right)$, the $S^1$ part of the conformal Killing equations with time derivative \eqref{time chemical} reduces to
\be
\p_0\e=-\frac{\tau-\beta}{2\tau+\beta}\left(1-i\gamma^3\gamma^0\right)\gamma^5\e~.
\label{another s1}
\ee 
Again, the projection operator $1-i\gamma^3\gamma^0$ allows only the generators for $Q$ and $S$ to be well-defined around the temporal circle $S^1$. Hence, the action with the time derivative \eqref{time chemical} has supersymmetries $Q$ and $S$ generated by \eqref{explicit kspinors2}. For $\tau=\beta$, the fermionic symmetries $Q^+_4$ and $S^4_+$ are restored.

The eigenvalues of the Laplacian $\nabla^2$ acting on divergenceless vector fields $B_j$ are $-(j+1)^2$, where $j$ is an integer $\ge 1$. Here the eigenfunctions (the spherical harmonics on $S^3$ with spin $1$) transform as the representation $(j_3,\overline j_3)=(\frac{j+1}2,\frac{j-1}2) \oplus(\frac{j+1}2,\frac{j-1}2)$ under the rotational group $SO(4)\cong SU(2)_L\times SU(2)_R$. In the representation  $(j_3,\overline j_3)=(\frac{j+1}2,\frac{j-1}2) $, the eigenvalues of $i\nabla_3$ and $i\overline\nabla_3$ runs $(-\frac{j+1}2,\cdots,\frac{j+1}2)$ and $(-\frac{j-1}2,\cdots, \frac{j-1}2)$ and hence they occur with degeneracy $j(j+2)$.  Thus the bosonic part of the determinant is:

\begin{footnotesize}
\begin{eqnarray}
  &&
 { \det}_{\rm vect}\left[-\left(\partial_0-i\frac{\alpha_{\rm adj}}
  {\beta\!+\!2\tau}+ \frac{2(\tau\!-\!\beta)}{\beta+2\tau}(i\nabla_3)
  +\frac{2\gamma}{\beta\!+\!2\tau}(i\overline \nabla_3) +\frac{\frac12\beta\!+\!\zeta_1}{\beta\!+\!2\tau} R_1
  +\frac{\beta\!+\!\zeta_2}{\beta+2\tau}R_2+\frac{\frac32\beta}{\beta\!+\!
  2\tau}R_3  \right)^2-\nabla^2\right]\cr
  &&\hspace{-0.3cm}=\!\prod_{n=-\infty}^\infty\!\prod_{j,j_3,\overline j_3}
  \left[\left(\frac{2\pi n}{\beta\!+\!2\tau}
  -\frac{\alpha_{\rm adj}}{\beta\!+\!2\tau}
 -i \frac{2(\tau\!-\!\beta)}{\beta+2\tau}j_3
  -i\frac{2\gamma}{\beta\!+\!2\tau}\overline j_3 -i\frac{\frac12\beta\!+\!\zeta_1}{\beta\!+\!2\tau} r_1
  -i\frac{\beta\!+\!\zeta_2}{\beta+2\tau}r_2-i\frac{\frac32\beta}{\beta\!+\!
  2\tau}r_3
  \right)^2\!+\!\left(j\!+\!1\right)^2\right] \cr
  &&
\end{eqnarray}
\end{footnotesize}

Following the prescription in \cite{Aharony:2003sx}, we factor
out a divergent constant, set it to unity, and obtain
{\small
\begin{eqnarray}\label{scalar-pair-det}
  {\det}_{\rm vect}^{-\frac12}&=&\prod_{j,j_3,\bar{j_3}}(-2i)
  \sin\left[\frac{1}{2}\left(\!\a_{\rm adj}+i\beta\Delta^{\!+}+2i\tau(\e_j^{(1)}\!+\!j_3)+i(2\gamma \overline j_3+\zeta_1r_1\!+\!\zeta_2r_2)
  \right)\right]\cr
  &&\hspace{3cm}\times(-2i)
  \sin\left[\frac{1}{2}\left(\!-\a_{\rm adj}+i\beta\Delta^{\!-}+2i\tau(\e_j^{(1)}\!-\!j_3)-i(2\gamma \overline j_3+\zeta_1r_1\!+\!\zeta_2r_2)
  \right)\right]\cr
 &=&\prod_{j,j_3,\overline j_3}
  e^{(\beta+2\tau)\e_j^{(1)}}  \left(1-e^{i\a_{\rm adj}}
  x^{\Delta^{\!+}} t^{2(\e_j^{(1)}\!+\!j_3)}y^{2\overline j_3}v^{r_1}w^{r_2}\right) \left(1-e^{-i\a_{\rm adj}}
  x^{\Delta^{\!-}} t^{2(\e_j^{(1)}\!-\!j_3)}y^{-2\overline j_3}v^{-r_1}w^{-r_2}\right) \cr
  && \label{one-loop}
\end{eqnarray}}
where $\e_j^{(1)}\equiv j+1$ and $\Delta^{\!\pm}\equiv\e_j\mp2j_3\!\pm\frac12 r_1\pm r_2\pm\frac32r_3$. Since we take only $\Delta^{\!\pm} \ge 0$, the expression \eqref{one-loop} is $ {\det}_{\rm vect}^{-\frac12}$ instead of $ {\det}_{\rm vect}$. To write $ {\det}_{\rm vect}^{-\frac12}$ in terms of 
single-particle index as in the $\N=4$ index \eqref{spsi}, we manipulate \eqref{one-loop} as 
{\small
\begin{eqnarray}\label{scalar-det-log}
\log({\det}_{\rm vect}^{-\frac12})
  &\equiv&-(\beta+2\tau)N^2\sum_{j=1}^\infty 2j(j+2)\e_j^{(1)}
 +\sum_{m=1}^\infty\frac{1}{m}\left[
  f^B_{\rm vect}(x^m,t^m,y^m,v^m,w^m) \text{Tr}(U^\dag)^m \text{Tr}(U)^m\}
  \!\frac{}{}\right]\ .\cr
  &&
\end{eqnarray}}
where $x\equiv e^{-\beta}$.
The first term provides a quantity analogous to the Casimir energy,
which was computed in \cite{Kinney:2005ej}. (See around (4.26) in  \cite{Kinney:2005ej}) The contribution from
the gauge field to the single-particle index is given by
\bea
  f^B_{\rm vect}(x,t,y,v,w)& \equiv&
  \sum_{j=1}^\infty\sum_{(j_3,\overline j_3)=(\frac{j+1}2,\frac{j-1}2)}
  \left(x^{\Delta^{\!+}} t^{2(\e_j^{(1)}\!+\!j_3)}y^{2\overline j_3}v^{r_1}w^{r_2}\right)\cr
  &+& \sum_{j=1}^\infty\sum_{(j_3,\overline j_3)=(\frac{j-1}2,\frac{j+1}2)} \left(x^{\Delta^{\!-}} t^{2(\e_j^{(1)}\!-\!j_3)}y^{-2\overline j_3}v^{-r_1}w^{-r_2}\right)
\eea
Explicitly summing over all the vector modes on $S^3$, one obtains
\begin{eqnarray}
  f^B_{\rm vect}(x,t,y,v,w)&=&
  \sum_{j=1}^\infty 
  \left[\left(\sum_{n=0}^{j-1}y^{j-1-2n}\right)\left(t^{3(j+1)}+t^{3j+1}x^2+\cdots+
  t^{j+3} x^{2j}+t^{j+1}x^{2(j+1)}\right)\right]\cr
  &+&   \sum_{j=1}^\infty 
  \left[\left(\sum_{n=0}^{j+1}y^{j+1-2n}\right)\left(t^{3j+1}x^2+t^{3j-1}x^4+\cdots+
  t^{j+5} x^{2j-2}+t^{j+3}x^{2j}\right)\right] \ .\cr
&&
\end{eqnarray}

Next, we consider the Pfaffian from the $\N=1$ gaugino. For the $\N=1$ gaugino, we note that, on $S^3$, the  eigenvalues of the Dirac operator $i \nslash$ acting on Weyl spinors are $\pm(j+\frac{1}{2})$ whose eigenfunctions (the spherical harmonics on $S^3$ with spin $\frac12$) transform as the representation $(j_3,\overline j_3)=(\frac j2,\frac {j-1}2)\oplus(\frac {j-1}2,\frac {j}2)$. ($j$ runs over the positive integers.) Analogous to the bosonic determinant, one can also write the Pfaffian of the Dirac operator $i~\nslash$ in terms of indices over letters
as follows:

{\footnotesize
\begin{eqnarray}\label{ferm-det-log}
  \log({\rm Pfaff}_{\rm vect}) &\equiv&+ (\beta+2\tau)N^2\sum_{j=1}^\infty 2j(j+1)\e^{(\frac12)}_j
-\sum_{m=1}^\infty\frac{1}{m}\left[
  f^F_{\rm vect}(x^m,t^m,y^m,v^m,w^m) \text{Tr}(U^\dag)^m \text{Tr}(U)^m\}
  \!\frac{}{}\right]\ , \cr
  &&
  \end{eqnarray}}
where $\e^{(\frac12)}_j\equiv j+\frac12$ and the single-particle index for the $\N=1$ gaugino is given by
\bea
 f^F_{\rm vect}(x,t,y,v,w)& \equiv&
  \sum_{j=1}^\infty\sum_{(j_3,\overline j_3)=(\frac{j}2,\frac{j-1}2)}
  \left(x^{\Delta^{\!+}} t^{2(\epsilon^{(\frac12)}_j\!+\!j_3)}y^{2\overline j_3}v^{r_1}w^{r_2}\right)\cr
  &+&  \sum_{j=1}^\infty\sum_{(j_3,\overline j_3)=(\frac{j-1}2,\frac{j}2)}
  \left(x^{\Delta^{\!-}} t^{2(\epsilon^{(\frac12)}_j\!-\!j_3)}y^{-2\overline j_3}v^{-r_1}w^{-r_2}\right) \ .
\eea
Note that we use the fact that the fermionic fields are periodic around the temporal circle $S^1$ here. Evaluating all the spinor modes on $S^3$, one obtains
\begin{eqnarray}
  f^F_{\rm vect}(x,t,y,v,w)&=&
   \sum_{j=1}^\infty 
  \left[\left(\sum_{n=0}^{j-1}y^{j-1-2n}\right)\left(t^{3j+1}x^2+t^{3j+3}x^4+\cdots+
  t^{j+3} x^{2j}+t^{j+1}x^{2(j+1)}\right)\right] \cr
    &+& \sum_{j=1}^\infty 
  \left[\left(\sum_{n=0}^{j}y^{j-2n}\right)\left(t^{3j}+t^{3j-2}x^2+\cdots+
  t^{j+4} x^{2j-4}+t^{j+2}x^{2j-2}\right)\right] \ .\cr
&&
\end{eqnarray}
Dropping the Casimir energies\footnote{We are not concerned with the Casimir energy here since it has already been argued in \cite{Balasubramanian:1999re} and \cite{Aharony:2003sx}. (See Eq.~(64) in \cite{Balasubramanian:1999re} and the footnote 30 in \cite{Aharony:2003sx}.)},  which are irrelevant to the $\N=4$ index, we combine the bosonic and fermionic determinants of the vector multiplet as
\begin{eqnarray}\label{matter-det-log}
\log\left(\frac{{\rm Pfaff}_{\rm vect}}
  {\sqrt{\det\!{}_{\rm vect}}}\right)&=&\sum_{m=1}^\infty\frac{1}{m}\left[
  f_{\rm vect}(x^m,t^m,y^m,v^m,w^m) \text{Tr}(U^\dag)^m \text{Tr}(U)^m\}
  \!\frac{}{}\right]
\end{eqnarray}
where the single-particle index $f_{\rm vect}$ of the vector multiplet takes a rather simple form due to the huge cancellation between bosonic and fermionic modes:
\bea
   f_{\rm vect}(x^m,t^m,y^m,v^m,w^m)&\equiv&f^B_{\rm vect}-f^F_{\rm vect}\cr&=&
  \frac{t^6}{(1-yt^3)(1-t^3/y)}
  -\left(1-\frac{1}{(1-yt^3)(1-t^3/y)}\right)\cr
  &=&\frac{-(y+y^{-1})t^3+2t^6}{(1-yt^3)(1-t^3/y)} \ .
\eea
We can see that  the single-particle index $f_{\rm vect}$ is independent of the circumference $\beta$ of the time circle $S^1$ since only the terms without the fugacity $x$, {\it i.e.} $\Delta=0$, survive as expected from the definition of the $\N=4$ index.

It is straightforward to compute the contributions from the chiral multiplets. With the time derivative \eqref{time chemical}, the one-loop determinant of the scalar fields $\phi^j$ can be written as $\det_{\rm scalar}[-D_0^2-\nabla^2+1]$ where the last constant comes from the curvature coupling term. Note that the coefficient $1$ of the curvature coupling term becomes important here to have nice square roots $\e_j^{(0)}\equiv j+1$
since the eigenvalues of the Laplacian $\nabla^2$ are $-j( j+2)$.
The eigenfunctions (the spherical harmonics on $S^3$ with spin $0$) transform as the representation $(j_3,\overline j_3)=(\frac j2,\frac j2)$. Thus, one can write the single-particle index for the scalar fields
\bea
  f^B_{\rm chiral}(x,t,y,v,w)& \equiv&
 \sum_{j=0}^\infty\sum_{(j_3,\overline j_3)=(\frac{j}2,\frac{j}2)}
  \left(x^{\Delta^{\!+}} t^{2(\epsilon^{(0)}_j\!+\!j_3)}y^{2\overline j_3}v^{r_1}w^{r_2}\right)\cr
  &+&  \sum_{j=0}^\infty\sum_{(j_3,\overline j_3)=(\frac{j}2,\frac{j}2)}
  \left(x^{\Delta^{\!-}} t^{2(\epsilon^{(0)}_j\!-\!j_3)}y^{-2\overline j_3}v^{-r_1}w^{-r_2}\right)
  \label{scalar}
\eea
 Enumeration over all the scalar modes gives us
\begin{eqnarray}
 && f^B_{\rm chiral}(x,t,y,v,w)\cr
 &=&
 \left(v+\frac 1w+\frac wv\right)   \sum_{j=0}^\infty 
  \left[\left(\sum_{n=0}^{j}y^{j-2n}\right)\left(t^{3j+2}x^2+t^{3j}x^4+\cdots+
  t^{j+4} x^{2j}+t^{j+2}x^{2j+2}\right)\right]\cr
  &+&  \left(w+\frac 1v+\frac vw\right) \sum_{j=0}^\infty 
  \left[\left(\sum_{n=0}^{j}y^{j-2n}\right)\left(t^{3j+2}+t^{3j}x^2+\cdots+
  t^{j+4} x^{2j-2}+t^{j+2}x^{2j}\right)\right] \ .\cr
&&
\end{eqnarray}
Similar to the $\N=1$ gaugino, we can write the single-particle index for the fermionic fields $\lambda^j$ as
\bea
 f^F_{\rm chiral}(x,t,y,v,w)& \equiv&
  \sum_{j=1}^\infty\sum_{(j_3,\overline j_3)=(\frac{j}2,\frac{j-1}2)}
  \left(x^{\Delta^{\!+}} t^{2(\epsilon^{(\frac12)}_j\!+\!j_3)}y^{2\overline j_3}v^{r_1}w^{r_2}\right)\cr
  &+&  \sum_{j=1}^\infty\sum_{(j_3,\overline j_3)=(\frac{j-1}2,\frac{j}2)}
  \left(x^{\Delta^{\!-}} t^{2(\epsilon^{(\frac12)}_j\!-\!j_3)}y^{-2\overline j_3}v^{-r_1}w^{-r_2}\right) \ .
\eea
We can write this more explicitly
\begin{eqnarray}
 && f^F_{\rm chiral}(x,t,y,v,w)\cr&=&
 \left(v+\frac 1w+\frac wv\right)  \sum_{j=1}^\infty 
  \left[\left(\sum_{n=0}^{j-1}y^{j-1-2n}\right)\left(t^{3j+1}+t^{3j-1}x^2+\cdots+
  t^{j+3} x^{2j-2}+t^{j+1}x^{2j}\right)\right] \cr
    &+&\left(w+\frac 1v+\frac vw\right) \sum_{j=1}^\infty 
  \left[\left(\sum_{n=0}^{j}y^{j-2n}\right)\left(t^{3j}x^2+t^{3j-2}x^4+\cdots+
  t^{j+4} x^{2j-2}+t^{j+2}x^{2j}\right)\right] \ .\cr
&&
\end{eqnarray}

Putting both the bosonic and fermionic pieces together, the one-loop determinant of the chiral multiplets can be casted up to Casimir energy in the following form
\begin{eqnarray}
 \log\left(\frac{{\rm Pfaff}_{\rm chiral}}
  {\det\!{}_{\rm chiral}}\right)&=&\sum_{m=1}^\infty\frac{1}{m}\left[
  f_{\rm chiral}(x^m,t^m,y^m,v^m,w^m) \text{Tr}(U^\dag)^m \text{Tr}(U)^m\}
  \!\frac{}{}\right]\ .
\end{eqnarray}
where all the terms with the fugacity $x$ again cancel between bosonic and fermionic modes
\bea
   f_{\rm chiral}(x^m,t^m,y^m,v^m,w^m)&\equiv&f^B_{\rm chiral}-f^F_{\rm chiral}\cr&=&
  \frac{t^2(w+\frac 1v+\frac vw)}{(1-yt^3)(1-t^3/y)}
  -\frac{t^4 (v+\frac 1w+\frac wv)}{(1-yt^3)(1-t^3/y)} \ .
\eea

All in all, we can write the one-loop determinants as
\bea
Z_{\rm 1-loop}=\exp \left\{ \sum_{m=1}^\infty \frac 1m
f(t^m,y^m,v^m,w^m) \text{Tr}(U^\dag)^m \text{Tr}\,  U^m\right\}
\eea
where the single-particle partition function $f(t,y,v,w) $ is a sum of the letter indices of the vector and chiral multiplets
\bea
f(t,y,v,w) &=&f_{\rm vect}(t,y,v,w)+f_{\rm chiral}(t,y,v,w)\cr
&=& \frac{t^2(v+\frac 1w + \frac wv) - t^3 (y+\frac 1y)
- t^4 (w+\frac 1v+\frac vw) + 2 t^6}{(1-t^3y)(1-\frac{t^3}{y})}
\eea 
Plugging this into \eqref{localizedpf}, we obtain the correct matrix integral for the $\N=4$ index as in \eqref{matrixintegral}.

\section{Conclusions and Future Directions}\label{section5}
In this paper, we interpret the $\N=4$ superconformal index as the partition function on the Scherk-Schwarz deformed background. We found the deformed action whose fermionic symmetries are only $Q$ and $S$ and generalize the action in the off-shell formulation to implement the localization methods. By writing the action as a $\delta_\e$-exact term where the conformal Killing spinor $\e$ generates $Q+S$, the partition function turns out to be localized at the space of the flat connections. We identify  the space of the flat connections on $S^1 \times S^3$ as the quotient space $T/W$, using  the fact that the flat connection can be classified the holonomy homomorphism. This also explains the reason why the Polyakov loop appears in the matrix integral form of the $\N=4$ index. The one-loop evaluations around the flat connections provides the correct single-particle index.

Finally, several technical and conceptual issues remain to be addressed
even within the direct line of attack of this paper. It is natural to generalize this functional integral interpretation to the $\N=1$ and $\N=2$ superconformal indices. Especially this will provide a rigorous explanation to the $\N=1$ index in which single-particle states are counted at UV. A large class of $\N=1$ SCFTs can be realized as strongly-coupled CFTs at IR fixed points whose UV theories are not conformal at quantum level in general. Applying the localization method to a UV theory, one may be able compute the partition function of the IR CFT exactly.

The other direction one may extend is the dimensional reduction of the partition function to three dimension as recently explored in \cite{Dolan:2011rp,Gadde:2011ia,Imamura:2011uw}. Following Nekrasov \cite{Nekrasov:2002qd}, the four-dimensional superconformal index reduces to a three-dimensional low energy effective theory as the size of the time circle shrinks to zero. This low energy effective field theory presumably contains all the information of the BPS states in the original four-dimensional SCFT. It was firstly shown in \cite{Dolan:2011rp} that starting from four-dimensional pair of Seiberg dual theories one can get the whole set of new dualities both for SYM and CS theories in three dimensions using some limits of identity for superconformal indices of four dimensional Seiberg dual field theories to partition functions of three dimensional dual field theories. Gereralizing the result of \cite{Gadde:2011ia}, it was also investigated in \cite{Imamura:2011uw} that three-dimensional partition functions with various parameters can be also obtained as a limit of the index of four-dimensional theories. Apart from these pioneering works, the feasibility of this approach still remains to be understood. This consideration is important since it might give new insights to BPS states in a SCFT with no Lagrangian description \cite{Gadde:2011ik}. For example, it is known that the compactification of the six-dimensional $(0,2)$ SCFT on a circle leads to the five-dimensional maximally supersymmetric Yang-Mills theory. It would be interesting to find a relation between the partition function of the  five-dimensional maximally SYM on $S^5$ and BPS states in the (0,2) theory. (The six-dimensional $(0,2)$ superconformal index in the large $N$ limit was computed from the gravity theory on $AdS_7\times S^4$ \cite{Bhattacharya:2008zy}.)

\section*{Acknowledgements} 
The author would like to express special gratitude to Xing Huang and Shiraz Minwalla for valuable discussions.
In addition, he would like to thank to Jyotirmoy Bhattacharya, Indranil Biswas, Giulio Bonelli, Atish Dabholkar, Abhijit Gadde, Rajesh Gopakumar, Suresh Govindarajan, Amihay Hanany, Kazuo Hosomichi, Seok Kim, Shailesh Lal, Marco Mari\~no, Jose F. Morales, Sameer Murthy, Kazumi Okuyama, Ramadas Ramakrishnan, Tarun Sharma, Alessandro Tanzini, Xi Yin and Jian Zhao for their helpful comments. He is also thankful to Massimo Bianchi, Leslaw Rachwal and Leonard Rastelli who provided me encouragement to complete this work. This research had been developed during ``School and Workshop on D-brane Instantons, Wall Crossing and Microstate Counting'' and ``School and Conference on Modular Forms and Mock Modular Forms and their Applications in Arithmetic, Geometry and Physics'' at the ICTP Trieste, and ``Indian Strings Meeting'' at Puri, India. Hence he is also gratetul for their stimulating academic environment and for their warm hospitality. 

\section*{Note added}
The author is grateful to the anonymous referee of JHEP
for careful reading of the manuscript and for raising important questions to improve the original version of this paper. In addition, he would also like to to Simone Giombi, Jaume Gomis, Robert Myers, Takuya Okuda, Wolfger Peelaers and Grigory Vartanov for helpful comments after this paper appeared on ArXiv.  
 He is also indebted to the Simons Summer Workshop in Mathematics and Physics 2011 for its stimulating academic environment and its warm hospitality since he benefited from discussions in the workshop.
 
\appendix

\section{Superconformal Algebra} \label{appa}
In this appendix, we review the four-dimensional ${\cal N} =m$ ($m=1,2$ or $4$) superconformal algebra in order for this paper to be self-contained and to establish notation. The notation is the same as in  \cite{Kinney:2005ej}.

In the Minkowski four dimensions $\R^{1,3}$, the
$SO(2,4)$ conformal algebra  is formed by the set of the  generators of
translations $P_{\mu}=- i \p_\mu$, of special conformal transformations  $K_\mu= i (2 x_\mu x \cdot \p -x^2
\p_\mu)$,
of the Lorentz group $SO(1,3)$, $M_{\mu \nu}= -i (x_\mu \p_\nu -x_\nu
\p_\mu)$,  and  of dilations $H=x\cdot \p$. The commutation relations have the form
\begin{equation}
\begin{array}{l}
[H,P_{\mu}] = P_{\mu}, \cr
 [M_{\mu \nu}, P_{\rho}] = i(\eta_{\mu \rho}
P_{\nu} - \eta_{\nu \rho} P_{\mu}) ,\cr
[M_{\mu \nu}, M_{\rho \sigma}] = i \left(\eta_{\mu \rho} M_{\nu
\sigma} + \eta_{\nu \sigma} M_{\mu \rho} - \eta_{\mu \sigma}
M_{\nu \rho} - \eta_{\nu \rho} M_{\mu \sigma} \right) \cr
\end{array}
\begin{array}{l}
 [H,K_{\mu}]= - K_{\mu} \cr
  [M_{\mu \nu},K_{\rho}] =
i (\eta_{\mu \rho} K_{\nu} - \eta_{\nu \rho} K_{\mu}), \cr
  [K_{\mu}, P_{\nu}] = 2 (\eta_{\mu \nu} H - i
M_{\mu \nu}) \cr
\end{array}\label{CA}
\end{equation}
where the metric $\eta_{\mu\nu}={\rm diag}(-,+,+,+)$  and the indices $\mu=0,1,2,3$. In terms of the matrix $M_{ab}$, we can put the algebra in more concise form;
\be
M_{ab} =        \left(
                \begin{array}{ccc}
                0 & H & \frac 12 (P_\nu-K_\nu) \\
                -H & 0 & \frac 12 (P_\nu+K_\nu) \\
                -\frac 12 (P_\mu-K_\mu) & -\frac 12 (P_\mu+K_\mu)  & M_{\mu\nu}
                \end{array}
                \right),
\ee
where we extend the indices to negative number, $a,b=-2,-1,0,\cdots,3$ and the commutation relations exhibits exactly the $SO(2,4)$ algebra;
\be\label{CA2}
[M_{ab},M_{cd}]=i(\eta_{ac}M_{bd}-\eta_{bc}M_{ad}-\eta_{ad}M_{bc}
+\eta_{bd}M_{ac}),\ee
with $\eta_{ab}={\rm diag}(-1,-1,1,1,1,1)$.
In the spinorial basis one defines
\begin{eqnarray}
 && P_{\alpha {\dot \alpha}} =
(\sigma^\mu)_{\alpha {\dot \alpha}}P_\mu, \ \ \ \ \ \ \ \ \ \ \ \ \ \ K^{{\dot \alpha}
\alpha}=(\overline{\sigma}^\mu)^{{\dot \alpha} \alpha}K_\mu, \cr 
&& J_{\alpha}^{~\beta}=\frac i4
(\sigma^\mu\overline{\sigma}^\nu)_{\alpha}^{~\beta}M_{\mu\nu}, \ \ \ \ \
\overline{J}^{{\dot \alpha}}_{~\dot\beta}=\frac i4
(\overline{\sigma}^\mu\sigma^\nu)^{{\dot \alpha}}_{~ \dot\beta}M_{\mu\nu}, 
\end{eqnarray}
 where
\be\sigma^\mu \ = \ (iI,i\sigma^i), \ \ \ \ \ \overline{\sigma}^\mu \ = \
(iI,-i\sigma^j)\label{foursigma}\ee and $\sigma^j$ are the Pauli matrices.
\begin{equation}
\sigma^1 \ = \ \left(
                        \begin{array}{cc}
                          0 & 1 \\
                          1 & 0 \\
                        \end{array}
                      \right), \ \ \
\sigma^2 \ = \ \left(
                        \begin{array}{cc}
                          0 & -i \\
                          i & 0 \\
                        \end{array}
                      \right), \ \ \
\sigma^3 \ = \  \left(
                        \begin{array}{cc}
                          1 & 0 \\
                          0 & -1 \\
                        \end{array}
                      \right). \label{pauli}
\end{equation}
The generators $J$ and $\overline J$ of $SU(2)_L\times SU(2)_R$ can be written by using the standard angular momentum generators, 
\be
J_{\alpha}^{\ \beta}=\left(\begin{array}{cc}
  J_3 & J_+ \\
  J_- & -J_3
\end{array}\right), \ \ \ \overline{J}_{\ \dot{\beta}}^{\dot{\alpha}}
=\left(\begin{array}{cc}
  \overline{J}_3 & \overline{J}_+ \\
  \overline{J}_- & -\overline{J}_3
\end{array}\right), \label{angmom} 
\ee
with 
\be
[J_+,J_-]=2J_3, \qquad
[\overline{J}_+,\overline{J}_-]=2\overline{J}_3.
\ee
Then the generators $M_{\mu\nu}$ of $SO(1,3)$ are expressed through these operators as
\bea
&&
{\scriptsize
M_{ab}=\left(\begin{array}{cccc}
  0 & \frac i2 (\overline{J}_+ + \overline{J}_- - J_+ - J_- ) & \frac 12 (J_+ + \overline{J}_- - \overline{J}_+ - J_-) & i (\overline{J}_3 - J_3)  \\
  -\frac i2 (\overline{J}_+ + \overline{J}_- - J_+ - J_- ) & 0 & -(J_3 + \overline{J}_3) & \frac i2 (J_+ + \overline{J}_+ - J_- - \overline{J}_-) \\
  -\frac 12 (J_+ + \overline{J}_- - \overline{J}_+ - J_-) & (J_3 + \overline{J}_3) & 0 & -\frac 12 (J_+ + J_- + \overline{J}_+ + \overline{J}_- ) \\
  -i (\overline{J}_3 - J_3) & -\frac i2 (J_+ + \overline{J}_+ - J_- - \overline{J}_-)
  & \frac 12 (J_+ + J_- + \overline{J}_+ + \overline{J}_- ) & 0
\end{array}\right). 
}
\cr 
&&
\eea
We rewrite the $SO(2,4)$ conformal algebra, \eqref{CA} and \eqref{CA2} in aal basis;
\begin{equation}
\begin{array}{l}
[J^{~\alpha}_{\beta}, J^{~\gamma}_{\delta}] =
\delta^{\gamma}_{\beta} J^{~\alpha}_{\delta} -
\delta^{\alpha}_{\delta} J^{~\gamma}_{\beta} \cr
[J^{~\alpha}_{\beta},P^{\gamma \dot{\delta}}] =
\delta^{\gamma}_{\beta} P^{\alpha \dot{\delta}} - \frac12
\delta^{\alpha}_{\beta} P^{\gamma \dot{\delta}}\cr
[\overline{J}^{\dot{\alpha}}_{~\dot{\beta}},P^{\dot{\delta} \gamma}] =
\delta^{\dot{\delta}}_{\dot{\beta}} P^{\dot{\alpha} \gamma} - \frac12
\delta^{\dot{\alpha}}_{\dot{\beta}} P^{\dot{\delta} \gamma}\cr
[H,P^{\alpha \dot{\beta}}] = P^{\alpha \dot{\beta}} \cr
 [K_{\alpha
\dot{\beta}}, P^{\gamma \dot{\delta}}] =
\delta^{\dot{\delta}}_{\dot{\beta}} J^{~\gamma}_{\alpha}+
\delta^{\gamma}_{\alpha}\overline{J}^{\dot{\delta}}_{~\dot{\beta}} +
\delta^{\dot{\delta}}_{\dot{\beta}} \delta^{\gamma}_{\alpha} H 
\end{array} \ \ \ \ \ \ \ \ \ \
\begin{array}{l}
[\overline{J}^{\dot{\alpha}}_{~\dot{\beta}},
\overline{J}^{\dot{\gamma}}_{~\dot{\delta}}] =
\delta^{\dot{\gamma}}_{\dot{\beta}}
\overline{J}^{\dot{\alpha}}_{~\dot{\delta}} -
\delta^{\dot{\alpha}}_{\dot{\delta}}
\overline{J}^{\dot{\gamma}}_{~\dot{\beta}} \cr
[J^{~\alpha}_{\beta},K_{\gamma \dot{\delta}}] =
\delta_{\gamma}^{\alpha} K_{\beta \dot{\delta}} - \frac12
\delta^{\alpha}_{\beta} K_{\gamma \dot{\delta}}\cr
[\overline{J}^{\dot{\alpha}}_{~\dot{\beta}},K_{\dot{\delta} \gamma}] =
\delta_{\dot{\delta}}^{\dot{\alpha}} K_{\dot{\beta} \gamma} - \frac12
\delta^{\dot{\alpha}}_{\dot{\beta}} K_{\dot{\delta} \gamma}\cr
[H,K^{\alpha \dot{\beta}}] = -K^{\alpha \dot{\beta}} 
\end{array}
\end{equation}
Now we extend the $SO(2,4)$ conformal algebra to the ${\cal N}=m$ superconformal algebra by introducing supercharges $Q^\alpha_{ A}, {\overline Q}_{\dot\alpha }^A$ and their superconformal partners $S_{\alpha }^A, {\overline S}^{\dot\alpha }_A$ with $A=1,\cdots,m$. The supercharges obey the anti-commutation relations 
\bea
&&\{Q^{\alpha}_A, {\overline Q}^{\dot \alpha B} \} = P^{\alpha {\dot \alpha}} \delta_A^B, \cr
&&\{ S_{\alpha }^A , {\overline S}_{\dot \alpha B}\} = K_{\alpha {\dot \alpha}} \delta^A_B,  \cr
&& \{S_{\alpha }^A,Q^{\beta }_B \} = \delta^{A}_{B} J^{~\beta}_{\alpha} +\delta^{\beta}_{\alpha} R^{A}_{B}+ \delta^{A}_{B}\delta^{\beta}_{\alpha} \left({H \over 2} + r{4 -m \over 4 m}\right) , \cr
  &&\{\overline S_{\dot\alpha A},\overline Q^{\dot\beta B} \} = \delta_{A}^{B} \overline J^{\dot\beta}_{~\dot\alpha} -\delta^{\dot\beta}_{\dot\alpha} R_{A}^{B} + \delta_{A}^{B}\delta^{\dot\beta}_{\dot\alpha} \left({H \over 2} - r{4 -m \over 4 m}\right)
\eea
and the other anti-commutation relations vanish.  Here $R_A^B$ and $r$ are the generators of $U(m)$ $R$-symmetry, except in the special case $m=4$, where the $R$-symmetry algebra is $SU(4)$. The commutation relations between bosonic and fermionic generators are listed in detail as follows;
\begin{equation}
\begin{array}{l}
[J^{~\alpha}_{\beta},Q^{\gamma}_A] =  \delta^{\gamma}_{\beta}
Q^{\alpha }_A -\frac12 \delta^{\alpha}_{\beta} Q^{\gamma }_A \cr
[K_{\alpha
\dot{\beta}},Q^{\gamma }_A] = \delta^{\gamma}_{\alpha}
\overline{S}_{\dot{\beta} A}, \cr
[H,Q^{\gamma }_A] = \frac{1}{2} Q^{\gamma }_A , \cr
[H, S_{\alpha }^A]= -\frac{1}{2} S_{\alpha }^A, \cr
[r,Q^{\gamma }_A]  = Q^{\gamma
}_A , \cr
 [r,S_{\alpha }^A] =- S_{\alpha }^A ,  \cr
  [R^{B}_{A}, Q^{\alpha }_C] =
\delta^{B}_{C} Q^{\alpha }_A - \frac1m \delta^{B}_{A} Q^{\alpha }_C \cr
\end{array} \ \ \ \ \ \ \ \ \ \ 
\begin{array}{l}
[\overline J^{\dot\alpha}_{~\dot\beta},\overline{Q}^{\dot\gamma A}] =
\delta^{\gamma}_{\beta} \overline{Q}^{\dot\alpha A} - \frac12
\delta^{\alpha}_{\beta} \overline{Q}^{\dot\gamma A} \cr 
 [P_{\alpha
\dot{\beta}},\overline{Q}^{\gamma A}] = \delta^{\gamma}_{\alpha} S^A_{
\dot{\beta}} \cr 
 [H,
{\overline Q}^{{\dot \alpha}A}]=\frac{1}{2} {\overline Q}^{{\dot \alpha}A},  \cr
 [H, {\overline S}_{{\dot
\alpha}A}] =-\frac{1}{2} {\overline S}_{{\dot \alpha}A} \cr
 [r, {\overline Q}^{{\dot \alpha}A}]=-{\overline Q}^{{\dot
\alpha}A}, \cr 
[r, {\overline
S}_{{\dot \alpha }A}] = {\overline S}_{{\dot \alpha }A} \cr
[R^{B}_{A}, R^{D}_{C}] = \delta^{B}_{C} R^{D}_{A} - \delta^{D}_{A}
R_{C}^{B} 
\end{array}
\end{equation}

Finally, we conclude this appendix by fixing the convention of the gamma matrices both in the Minkowski and Euclidean signature. In the Minkowski signature, we already specified the form of the gamma matrices in \eqref{foursigma}:
\bea
\gamma^\mu= \left(
                        \begin{array}{cc}
                          0 & \sigma^\mu \\
                          \bar\sigma^\mu & 0 \\
                        \end{array}
                      \right), \ \ \
\gamma^0= \left(
                        \begin{array}{cc}
                          0 &- i \\
                        - i & 0 \\
                        \end{array}
                      \right), \ \ \          
\gamma^j= \left(
                        \begin{array}{cc}
                          0 & i\sigma^j \\
                         -i\sigma^j & 0 \\
                        \end{array}
                      \right). \ \ \                                  
\eea
We define the chirality operator $\gamma^5= i\gamma^0\gamma^1\gamma^2\gamma^3$. The Euclidean case easily follows from the Wick rotation $x^0_E=ix^0$.
\bea
\gamma_E^\mu= \left(
                        \begin{array}{cc}
                          0 & \sigma_E^\mu \\
                          \bar\sigma_E^\mu & 0 \\
                        \end{array}
                      \right), \ \ \
\gamma_E^0= \left(
                        \begin{array}{cc}
                          0 & 1 \\
                         1 & 0 \\
                        \end{array}
                      \right), \ \ \          
\gamma_E^j= \left(
                        \begin{array}{cc}
                          0 & i\sigma^j \\
                         -i\sigma^j & 0 \\
                        \end{array}
                      \right). \ \ \          
                      \label{gamma euclid}                        
\eea
where we define 
\be
\sigma_E^\mu \ = \ (I,i\sigma^j), \ \ \ \ \ \overline{\sigma}_E^\mu \ = \
(I,-i\sigma^j)
\label{gam}
\ee
The chirality operator is $\gamma_E^5= \gamma_E^0\gamma_E^1\gamma_E^2\gamma_E^3$. 
Elsewhere, we omit the lower index $E$ for simplicity.

\section{Notations for Field Theory on $\R \times S^3$} \label{appb}
In this appendix, we explain how we obtain the four-dimensional $\N=4$ SCFT \eqref{actionSU(4)symmetric} from the ten-dimensional $\N=1$ SCFT \eqref{action}. We refer the reader to \cite{Okuyama:2002zn,Ishiki:2006rt} for more details.
\begin{eqnarray}
{\cal S}&=&\frac{1}{g_{YM}^2}
\int d^{10} x {\sqrt g}\; {\rm Tr}\left[\frac{1}{4} F_{MN}^2+\frac i2\bar{\lambda}\Gamma^MD_M\lambda+\frac{1}{12} RX_m^2\right]\cr
&=&\frac{1}{g_{YM}^2}\int d^4x {\sqrt g} \;  {\rm Tr}\Big[
\frac{1}{4} F_{\mu\nu}^2+\frac{1}{2}(D_{\mu}X_m)^2+  \frac{i}{2} \bar{\lambda}\Gamma^\mu D_\mu\lambda \cr
&& \hspace{4cm} +\frac{1}{2}\bar{\lambda}\Gamma^m[X_m,\lambda]+\frac{1}{4}[X_m,X_n]^2+\frac{1}{12}RX_m^2\Big]
\end{eqnarray}
The covariant derivative $D_\mu$ in the action \eqref{action} and \eqref{actionSU(4)symmetric} contains the spin connection $\omega^a_{~b}$ when it acts on the spinor fields.
 The spin connection $\omega^a_{~b} \in \Omega^1(End(T^*S^3))$   is non other than a uniquely determined natural connection, which sometimes called a Levi-Civita connection, satisfying
\bea
&&\omega^a_{~b}=-\omega^b_{~a} \cr
&&d e^a+\omega^a_{~b}\wedge  e^b=0 \ .
\eea 
where $\{e^a\}, \ a=0,\cdots 3$  is a vierbein on $\R \times S^3$.
Then the covariant derivative acting on a spinor (fermionic) field $\lambda$ can be written by using the spin connection $\omega^a_b$
\be
\nabla_\mu \lambda = \partial_\mu \lambda +\frac14 \omega_\mu^{ab} \Gamma_{ab} \lambda \ .
\ee
where $\omega_{\mu b}^a = \omega^{a}_{~b}(\partial_\mu) \in \Gamma(End(T^*S^3))$ and $\Gamma_{ab}$ is a Clifford multiplication of the gamma matrices. 
The action \eqref{action} is invariant under the superconformal transformation. In ten-dimensional notation, the transformation law is written as 
\bea
\delta_\e A_M=i\bar{\lambda}\Gamma_M\e,\ \ \ \ 
\delta_\e\lambda=\frac12 F_{MN}\Gamma^{MN}\e-\frac12 X_m\Gamma^m \Gamma^{\mu}\nabla_{\mu}\e \ .
\label{susytrans3}
\eea
Using four-dimensional Dirac spinor notation, it is also written as
\bea
&&\delta_{\epsilon}A_\mu=i\bar{\lambda}\Gamma_\mu\epsilon,\;\;\;
\delta_{\epsilon}X_m=i\bar{\lambda}\Gamma_m\epsilon, \cr
&&\delta_{\epsilon}\lambda=\left[\frac{1}{2}F_{\mu\nu}\Gamma^{\mu\nu}
+D_\mu X_m\Gamma^\mu\Gamma^m-\frac{1}{2}X_m\Gamma^m\Gamma^\mu\nabla_\mu
-\frac{i}{2}[X_m,X_n]\Gamma^{mn}\right]\epsilon \ .
\label{susytrans4}
\eea

The ten-dimensional Lorentz group has been decomposed as
$SO(1,9) \supset SO(1,3)\times SO(6)$. We identify $SO(6)$ with $SU(4)$.
We use $A,B=1,2,3,4$ as the indices of $\mbox{\boldmath $4$}$
in $SU(4)$ while we have used $m,n=4,\cdots,9$ as the indices of
$\mbox{\boldmath $6$}$ in $SO(6)$. The $SO(6)$ vector, $\mbox{\boldmath $6$}$,
corresponds to the antisymmetric tensor of $\mbox{\boldmath $4$}$ in $SU(4)$.
The $SO(6)$ and $SU(4)$ basis are related as
\bea
&&X_{i4}=\frac{1}{2}(X_{i+3}-iX_{i+6}) \;\;\; (i=1,2,3) \ , \cr
&&X_{AB}=-X_{BA} \ ,\;\;\;X^{AB}=-X^{BA}=X_{AB}^{\dagger} \ ,\;\;\;
X^{AB}=\frac{1}{2}\epsilon^{ABCD}X_{CD} \ ,
\label{sixdimension}
\eea
Similar identities hold for the gamma matrices:
\bea
\Gamma^{i4}=\frac{1}{2}(\Gamma^{i+3}+i\Gamma^{i+6}) \ ,
\label{b4}
\eea
The ten-dimensional gamma matrices are decomposed as
\bea
\Gamma^\mu=\gamma^\mu\otimes 1_8,\;\;\;
\Gamma^{AB}=\gamma_5\otimes \left( \begin{array}{cc}
                                   0          &  -\tilde{\rho}^{AB} \\
                                   \rho^{AB}  &  0
                                   \end{array}  \right)
=-\Gamma^{BA} \ ,
\eea
where $\gamma^\mu$ is the four-dimensional gamma matrix, satisfying
$\{\gamma^\mu,\gamma^\nu\}=2\eta^{\mu\nu}$, and 
$\gamma^5=i\gamma^0\gamma^1\gamma^2\gamma^3$. $\Gamma^{AB}$ satisfies
$\{\Gamma^{AB},\Gamma^{CD}\}=\epsilon^{ABCD}$, and $\rho^{AB}$ and
$\tilde{\rho}^{AB}$ are defined by
\bea
(\rho^{AB})_{CD}=\delta^A_C\delta^B_D-\delta^A_D\delta^B_C,\;\;\;
(\tilde{\rho}^{AB})^{CD}=\epsilon^{ABCD} \ .
\eea
The charge conjugation matrix and the chirality matrix are given by
\bea
C_{10}=C_4 \otimes \left( \begin{array}{cc}
                          0   &  1_4 \\
                          1_4 &  0   
                          \end{array} \right), \;\;\;\;\;
\Gamma^{11}=\Gamma^0\cdots\Gamma^9=\gamma_5\otimes 
            \left( \begin{array}{cc}
                     1_4   &  0    \\
                     0     &  -1_4    
                   \end{array} \right) \ ,
\eea
where $(\Gamma^{a,m})^T=-C_{10}^{-1}\Gamma^{a,m}C_{10}$ and 
$C_4$ is the charge conjugation matrix in four dimensions.
The Majorana-Weyl spinor in ten dimensions is decomposed as
\bea
\lambda=\Gamma_{11}\lambda
=\left(\begin{array}{c}{\lambda_\uparrow}^A \\ \lambda_{\downarrow A} \end{array}\right) \ ,
\label{10to4}
\eea
where $\lambda_{\downarrow A}$ is the charge conjugation of ${\lambda_\uparrow}^A$:
\bea
\lambda_{\downarrow A}=({\lambda_\uparrow}^A)^c=C_4(\overline{\lambda_\uparrow}_{ A})^T,\;\;\;\;\;
\gamma_5\lambda_{\updownarrow}=\pm\lambda_{\updownarrow} \ ,  \ \ \ \ \overline{\lambda_\updownarrow}\gamma_5=\mp\overline{\lambda_\updownarrow} \  .
\label{chargeconj}
\eea
For the sake of convenience, we redefine the field contents
\be
X^{i4} \equiv\frac12\phi^i \ , \ \ X_{i4} \equiv\frac12\bar\phi_i \ , \ \ \lambda^{\;\; 4}_{\updownarrow} \equiv\chi_{\updownarrow} \ ,  \ \ \ \ \overline{\lambda_{\updownarrow}}_4 \equiv\overline{\chi_{\updownarrow}} \ , 
\label{redefinition}
\ee
where $X^{AB}$ can be expressed as a matrix form
\be 
X^{AB}=\frac{1}{2}\left(\begin{array}{cccc}
0 & \bar{\phi_{3}} &- \bar{\phi_{2}} & \phi^{1}\\
-\bar{\phi_{3}} & 0 & \bar{\phi_{1}} & \phi^{2}\\
\bar{\phi_{2}} & -\bar{\phi_{1}} & 0 & \phi^{3}\\
-\phi^{1} & -\phi^{2} & -\phi^{3} & 0\end{array}\right)  \ \ \ \ 
X_{AB}=\frac{1}{2}\left(\begin{array}{cccc}
0 & \phi^{3} & -\phi^{2} &\bar{\phi}_{1}\\
-\phi^{3} & 0 & \phi^{1} & \bar{\phi}_{2}\\
\phi^{2} & -\phi^{1} & 0 & \bar{\phi}_{3}\\
-\bar{\phi}_{1} &- \bar{\phi}_{2} &- \bar{\phi}_{3} & 0\end{array}\right) \ .
\ee
The action is rewritten in terms of the $SU(4)$ symmetric notation as follows:
\bea
{\cal S}&=&\frac{1}{g_{YM}^2}\int d^4x{\sqrt g} \;  {\rm Tr}\Big(
\frac{1}{4}F_{\mu\nu}F^{\mu\nu}+\frac{1}{2}D_\mu X_{AB}D^\mu X^{AB}
+i\overline{{\lambda}_{\uparrow}}_{ A}\gamma^{\mu}D_{\mu}{\lambda_\uparrow}^A+\frac{1}{2}X_{AB}X^{AB} \cr
&+&\overline{{\lambda}_{\uparrow}}_{ A}[X^{AB},\lambda_{\downarrow B}]+\overline{{\lambda}_{\downarrow}}^A[X_{AB},{\lambda_\uparrow}^B]
+\frac{1}{4}[X_{AB},X_{CD}][X^{AB},X^{CD}]\Big) \ . \cr
&&
\eea
The superconformal transformation \eqref{susytrans4} can also be rewritten in terms of the $SU(4)$ symmetric notation as in \eqref{susytrans}.

\section{Superconformal Transformations by $Q$ and $S$}\label{appc}
In this appendix, we shall write the superconformal transformations of the $\N=4$ field contents by $Q$ and $S$. From the transformations of the fermionic fields, we find the moduli space of the 1/16 BPS states.

It follows from \eqref{explicit kspinors2} that the superconformal transformation by $Q$ can be obtained by taking only $\e_-^4$ and all the other components of the conformal Killing spinors zero in \eqref{susytrans2}. 
 \begin{eqnarray}
\begin{array}{l}
[Q,A_{+\dot\alpha}] = 0\cr
[Q,A_{-\dot\alpha}] = -2i\overline{ \chi_\uparrow}_{\dot\a}  \cr
[Q,(F^{+})_+^{~+}]=-[Q,(F^{+})_-^{~-}]= -iD_{+\dot\a}\overline{ \chi_\uparrow}^{\dot\a} \cr
[Q,(F^{+})_-^{ ~  +}]=-2iD_{-\dot\a}\overline{ \chi_\uparrow}^{\dot\a} \cr
[Q,(F^{+})_+^{ ~  -}]=0 \cr
[Q,\phi^{j}]=2i{\lambda_\uparrow}_+^j \cr
[Q,\bar\phi_{j}]=0 \cr
\end{array} \ \ \ \ \ \ 
\begin{array}{l}
\{Q,\chi_{\uparrow+}\}=(F^{+})_+^{ ~  -}\cr
\{Q,\chi_{\uparrow-}\}=(F^{+})_-^{~-}-\frac i2[\phi^{j},\bar\phi_{j}]\cr
\{Q,{\chi_\downarrow}^{\dot\a}\}= 0\cr
\{Q,{\lambda_\uparrow}^j_+\}=0\cr
\{Q,{\lambda_\uparrow}^j_-\}=-\frac i2\epsilon^{jkl}[\bar\phi_{k},\bar\phi_{l}]\cr
\{Q,{\lambda_{\downarrow}}_j^{\dot\a}\}=[D^{\dot\a-}+(\bar\sigma^0)^{\dot\a-}]\bar\phi_{j}\cr
\end{array}
\label{BRSTQ}
\end{eqnarray}
where, in the last anti-commutation relation, the covariant derivative is $D^{\dot\a-}\bar\phi_j\\=\left[(\bar\sigma^0)^{\dot\a-} (D_0-2iJ_3-1)+(\bar\sigma^j)^{\dot\a-}D_j \right]\bar\phi_j$ as in \eqref{covariant derivative} and the term $(\bar\sigma^0)^{\dot\a-}\bar\phi_{j}$ comes from the $\e$-derivative terms. Using the equation of motion $\partial_{-\dot\a}\chi^{\;\;\dot\a}_{\downarrow}=0$ for the $\N=1$ guagino, we can verify that  $\bar{\phi}_j$, $\chi^{\;\;\dot\a}_{\downarrow}$, $\lambda^{\;\;j}_{\uparrow-}$ and $(F^+)_-^{~+}$ are elements of the $Q$-cohomology groups at zero coupling as mentioned in the subsection \eqref{SCI} since the gauge field $A_{\a\dot\a}$ is not  a gauge invariant operator and we can ignore quadratic terms such as $[\bar\phi_{k},\bar\phi_{l}]$ at zero coupling. From the BRST-like transformations \eqref{BRSTQ}, one can see that the field configurations annihilated by $Q$ is defined by the equations
\bea\left\{
\begin{array}{l}
s_1=(F^{+})_+^{ ~  -}=0\cr
s_2=(F^{+})_-^{~-}-\frac i2[\phi^{j},\bar\phi_{j}]=0 \cr
s_3=[D^{\dot\a-}+(\bar\sigma^0)^{\dot\a-}]\bar\phi_{j}=0  \cr 
s_4=\epsilon^{jkl}[\bar\phi_{k},\bar\phi_{l}]=0 \cr
\end{array}\right .
\label{modulieqns}
\eea
(See the section 5 in \cite{Witten:1991zz} for the reason.) It turns out that the equations \eqref{modulieqns} become essentially the same as \eqref{moduli1} and \eqref{moduli2} after the Wick rotation. (See Appendix \ref{appd} as a reference for the self-dual gauge field.)

We can see some of the main properties of the moduli space defined by \eqref{modulieqns} by means of vanishing theorems. 
\bea
{\cal S}_{\rm bosonic}&=&\int_M d^4x\sqrt g \left[\frac 14|s_1|^2+\frac14|s_2|^2+\frac14|s_3|^2+\frac18|s_4|^2\right]\cr
&=&\int_M d^4\sqrt g \left[\frac 14|(F^+)_-^{~+}|^2+\frac14|(F^{+})_-^{~-}-\frac i2[\phi^{j},\bar\phi_{j}]|^2\right.\cr&&\hspace{4cm}+\frac14\left|\left(D^{\dot\a-}+(\bar\sigma^0)^{\dot\a-}\right)\bar\phi_{j}\right|^2+\frac18\sum_{i,j=1}^3\left| [\phi^{i},\phi^{j}] \right|^2\Big]\cr
&=&\int_M d^4\sqrt g \left[\frac14|F^+|^2+\frac14D_\mu\phi^jD^\mu\bar\phi_j\right.\cr&&\hspace{4cm}+\frac18\left| [\phi^{j},\bar\phi_{j}] \right|^2+\frac18\sum_{i,j=1}^3\left| [\phi^{i},\phi^{j}] \right|^2+\frac1{16}R\phi^{j}\bar\phi_{j}\Big] \ ,\cr
&&
\eea
where  we used  the Weitzenb\"ock formula
\bea
D_{\a\dot\a}D^{\dot\a\b}=\frac12 \delta_\a^\b D_{\gamma\dot\a} D^{\dot\a\gamma}+\frac14 \delta_\a^\b R+F^{+\b}_{~\a} \ .
\eea
Since the Ricci scalar curvature $R$ of $M$ is a positive constant, the solution of \eqref{modulieqns} must have $\phi^{j}=0$ and $F^+=0$.

In similar fashion, we can show the superconformal transformation by $S$ by neglecting all the conformal Killing spinors except $\bar\e^{\dot+}_4$.
\begin{eqnarray}
\begin{array}{l}
[S,A_{\alpha\dot+}] = 0\cr
[S,A_{\alpha \dot-}] = 2i{\chi_\uparrow}_\a\cr
[S,(F^{-})^{\dot+}_{~\dot+}]=-[S,(F^{-})^{\dot-}_{~\dot-}]=-iD^{\dot-\a} {\chi_\uparrow}_\a \cr
[S,(F^{-})^{\dot+}_{~\dot-}]=2iD^{\dot+\a}  {\chi_\uparrow}_\a \cr
[S,(F^{-})^{\dot-}_{~\dot+}]=0 \cr
[S,\phi^{j}]=0 \cr
[S,\bar\phi_{j}]=2i\overline{\lambda_\uparrow}_{j\dot+}\cr
\end{array}  \ \ \ \ \ \ 
\begin{array}{l}
\{S,\chi_{\uparrow\a}\}=0\cr
\{S,{\chi_\downarrow}^{\dot+}\}=(F^-)^{\dot+}_{~ \dot+}+\frac i2[\phi^{j},\bar\phi_{j}]\cr
\{S,{\chi_\downarrow}^{\dot-}\}= (F^-)^{\dot-}_{~\dot+}\cr
\{S,{\lambda_\uparrow}^j_\a\}=[D_{\a\dot+}-(\sigma^0)_{\a\dot+}]\phi^{j}\cr
\{S,{\lambda_{\downarrow}}_j^{\dot+}\}=-\frac i2\epsilon_{jkl}[\phi^{k},\phi^{l}]  \cr
\{S,{\lambda_{\downarrow}}_j^{\dot-}\}=0 \cr
\end{array}
\end{eqnarray}
As before, the vanishing theorem shows that the field configurations annihilated by $S$ obey $F^-=0$ and $\phi^{j}=0$. Hence, a solution for the moduli space of the 1/16 BPS states is a flat connection $F=0$ with $\phi^{j}=0$ as found in the subsection \ref{criticalpt}.

\section{Self-dual and Anti-self-dual Field Strength in Euclidean Signature}\label{appd}
In this appendix, we consider self-dual and anti-self-dual two forms in the four-dimensional Euclidean space. 

The multiplication rules of $\sigma^\mu, \bar\sigma^\mu$ in \eqref{gam} can be expressed as
\begin{equation}
\label{sigmaMultiplication}
\sigma_\mu \bar{\sigma}_\nu= \delta_{\mu\nu} {\bf 1}_2 + i\sigma_j \eta^j_{\mu\nu}  , \ \ \ \  \bar{\sigma}_\mu \sigma_\nu = \delta_{\mu\nu}{\bf 1}_2 + i \sigma_j \bar{\eta}^j_{\mu\nu} .
\end{equation}
where $\eta^j_{\mu\nu}$ and  $\bar{\eta}^j_{\mu\nu}$ are called the self-dual 't Hooft eta symbols and  anti-self-dual 't Hooft eta symbols, respectively. The name stems from the fact that the eta symbols connect the $(\bf{3,1})$ and $(\bf{1,3})$ representations of $SU(2)_L\times SU(2)_R$ with the self-dual  and anti-self-dual two-form in four dimensions
\begin{equation}
\label{tHooftProjector}
\begin{aligned}
(\sigma_{\mu\nu})_\alpha^{~\beta}&\equiv \frac{1}{4} \left( \sigma_{\mu}\bar{\sigma}_\nu - \sigma_{\nu}\bar{\sigma}_\mu \right) = \frac{i}{2}\sigma_{i,\a}{}^\b \eta^i_{\mu\nu} \ , \\
(\bar\sigma_{\mu\nu})_{~\dot\beta}^{\dot\alpha} &\equiv \frac{1}{4}\left(\bar{\sigma}_{\mu}\sigma_{\nu} - \bar{\sigma}_{\nu}\sigma_{\mu}\right) = \frac{i}{2}\sigma_{i}{}^{\dot{\a}}{}_{\dot{\b}} \bar{\eta}^i_{\mu\nu}  \ .
\end{aligned} 
\end{equation}
The eta symbols can be represented by six $4\times 4$ matrices as follows:
\begin{equation}
\label{tHooftSymbol}
\begin{aligned}
\eta^1_{\mu\nu} &= \left(
\begin{array}{cccc}
0 & 1 & 0 & 0 \\
-1 & 0 & 0 & 0 \\
0 & 0 & 0 & 1 \\
0 & 0 & -1 & 0 
\end{array}
\right), & 
\eta^2_{\mu\nu} &= \left(
\begin{array}{cccc}
0 & 0 & 1 & 0 \\
0 & 0 & 0 & -1 \\
-1 & 0 & 0 & 0 \\
0 & 1 & 0 & 0 
\end{array}
\right), &
\eta^3_{\mu\nu} &= \left(
\begin{array}{cccc}
0 & 0 & 0 & 1 \\
0 & 0 & 1 & 0 \\
0 & -1 & 0 & 0 \\
-1 & 0 & 0 & 0 
\end{array}
\right),
\end{aligned}
\end{equation}
and
\begin{equation}
\label{anti-tHooftSymbol} 
\begin{aligned}
\bar{\eta}^1_{\mu\nu} &= \left(
\begin{array}{cccc}
0 & -1 & 0 & 0 \\
1 & 0 & 0 & 0 \\
0 & 0 & 0 & 1 \\
0 & 0 & -1 & 0 
\end{array}
\right), &
\bar{\eta}^2_{\mu\nu} &= \left(
\begin{array}{cccc}
0 & 0 & -1 & 0 \\
0 & 0 & 0 & -1 \\
1 & 0 & 0 & 0 \\
0 & 1 & 0 & 0 
\end{array}
\right), &
\bar{\eta}^3_{\mu\nu} &= \left(
\begin{array}{cccc}
0 & 0 & 0 & -1 \\
0 & 0 & 1 & 0 \\
0 & -1 & 0 & 0 \\
1 & 0 & 0 & 0 
\end{array}
\right). 
\end{aligned}
\end{equation}
One can check that the eta symbols satisfy the multiplication rules:
\bea
\eta^i \eta^j = - \delta^{ij}{\bf 1}_4 - \epsilon^{ijk}\eta^k,  \ \ \ \  \ \ \ 
\bar{\eta}^i \bar{\eta}^j = - \delta^{ij}{\bf 1}_4 - \epsilon^{ijk}\bar{\eta}^k .
\label{tHooftMultiplication}
\eea
Then it easily follows that they obey the Lie algebra $\mathfrak{su(2)}$.
\bea
\left[-i\frac{\eta^i}{2},-i\frac{\eta^j}{2}\right] =i \epsilon^{ijk} \left( -i\frac{\eta^k}{2} \right), \ \ \ 
\left[-i\frac{\bar{\eta}^i}{2},-i\frac{\bar{\eta}^j}{2}\right] =i \epsilon^{ijk} \left( -i\frac{\bar{\eta}^k}{2} \right).
\eea
In addition, they hold the relations
\bea
&&\bar \eta^j_{\mu\nu}=\eta^j_{\mu\nu}=\e_{j\mu\nu} \ \ \ \  \mu, \nu=1,2,3 \cr
&&\bar \eta^j_{\mu0}=\eta^j_{0\mu}=\delta_{j\mu} \ \ \ \ \ \mu=1,2,3 \cr
&& \eta^j_{\mu\nu}=-\eta^j_{\nu\mu} \cr
&& \bar \eta^j_{\mu\nu}=-\eta^j_{\nu\mu} \ .
\eea
Using the matrix forms of the eta symbols \eqref{tHooftSymbol} and \eqref{anti-tHooftSymbol}, the self-dual part $F_{\ \alpha}^{+\beta}$  of the gauge field strength can be written explicitly
\bea
 F_{\ \alpha}^{+\beta}\equiv F_{\mu\nu}(\sigma^{\mu\nu})_\alpha^{~\beta}&=& \left(\begin{array}{cc}i(F_{03}+F_{12})&
F_{02}+F_{31}+i(F_{01}+F_{23})\\
               -F_{02}-F_{31}+i(F_{01}+F_{23})&- i(F_{03}+F_{12})\end{array}
\right), \cr
F^+_{\alpha \beta}&=&\left(\begin{array}{cc}
F_{02}+F_{31}+i(F_{01}+F_{23})&- i(F_{03}+F_{12})\\
            - i(F_{03}+F_{12})&F_{02}+F_{31}-i(F_{01}+F_{23}) \end{array} \right) \ . 
            \label{SDFS}
\eea
Similarly, we have the form of the anti-self-dual part $F^{-\dot\alpha}_{\ \ \ \dot\beta}$ of the gauge field strength
\bea
F^{-\dot\alpha}_{\ \ \ \dot\beta}\equiv F_{\mu\nu}(\bar\sigma^{\mu\nu})^{\dot\alpha}_{~\dot\beta}&=&\left(\begin{array}{cc}i(-F_{03}+F_{12})&
-F_{02}+F_{31}+i(-F_{01}+F_{23})\\
               F_{02}-F_{31}+i(-F_{01}+F_{23})&i(F_{03}-F_{12})\end{array}
\right), \cr
F^{-\dot\alpha \dot\beta}&=&\left(\begin{array}{cc}
F_{02}-F_{31}+i(F_{01}-F_{23})&i(-F_{03}+F_{12})\\
            i(-F_{03}+F_{12})&F_{02}-F_{31}+i(-F_{01}+F_{23}) \end{array} \right) \ . 
\eea

\bibliography{Localization}{}
\bibliographystyle{utphys}

\end{document}